%% file: main.tex
\documentclass[11pt,english]{report}
\usepackage[T1]{fontenc}
\usepackage[latin9]{inputenc}
\usepackage[a4paper]{geometry}
\usepackage{fancyhdr}
\usepackage{color}
\usepackage{babel}
\usepackage{array}
\usepackage{float}
\usepackage{multirow}
\usepackage{amsmath}
\usepackage{graphicx}
\usepackage{setspace}
\usepackage{nomencl}
\providecommand{\printnomenclature}{\printglossary}
\providecommand{\makenomenclature}{\makeglossary}
\makenomenclature
\usepackage[unicode=true,pdfusetitle,
 bookmarks=true,bookmarksnumbered=false,bookmarksopen=false,
 breaklinks=false,pdfborder={0 0 0},pdfborderstyle={},backref=false,colorlinks=false]
 {hyperref}

\makeatletter

\newcommand{\noun}[1]{\textsc{#1}}
\providecommand{\tabularnewline}{\\}

\date{}
\usepackage{apacite}
\usepackage{caption}
\usepackage{mathptmx}
\usepackage{regexpatch}
\usepackage{sectsty}
\chapternumberfont{\Huge} 
\chaptertitlefont{\huge}
\renewcommand{\headrulewidth}{0pt}

\emergencystretch=\maxdimen
\newcommand\blfootnote[1]{%
  \begingroup
  \renewcommand\thefootnote{}\footnote{#1}%
  \addtocounter{footnote}{-1}%
  \endgroup
}

\@ifundefined{showcaptionsetup}{}{%
 \PassOptionsToPackage{caption=false}{subfig}}
\usepackage{subfig}
\makeatother

\begin{document}
\include{front}

\settowidth{\nomlabelwidth}{\textbf{TF.IDF}}
\printnomenclature{}

\nomenclature{\textbf{AMT}}{Amazon Mechanical Turk}

\nomenclature{\textbf{TGM}}{Transaction Graph Mining}

\nomenclature{\textbf{SGM}}{Single Graph Mining}

\nomenclature{\textbf{DM}}{Data Mining}

\nomenclature{\textbf{IR}}{Information Retrieval}

\nomenclature{\textbf{ML}}{Machine Learning}

\nomenclature{\textbf{TM}}{Text Mining}

\nomenclature{\textbf{NB}}{Naive Bayes}

\nomenclature{\textbf{LR}}{Logistic Regression}

\nomenclature{\textbf{KNN}}{K-Nearest Neighbor}

\nomenclature{\textbf{DT}}{Decision Tree}

\nomenclature{\textbf{NN}}{Neural Network}

\nomenclature{\textbf{SVM}}{Support Vector Machine}

\nomenclature{\textbf{RF}}{Random Forest}

\nomenclature{\textbf{AUC}}{Area Under ROC Curve}

\nomenclature{\textbf{POS}}{Parts of Speech}

\nomenclature{\textbf{LIWC}}{ Linguistic Inquiry and Word Count}

\nomenclature{\textbf{PU}}{Positive Unlabelled}

\nomenclature{\textbf{IDT}}{Interpersonal Deception Theory}

\nomenclature{\textbf{TF.IDF}}{Term Frequency- Inverse Document Frequency}

\nomenclature{\textbf{RCS}}{Reviewer Content Similarity}

\nomenclature{\textbf{CSV}}{Comma Separated Values}

\chapter{Introduction}

\thispagestyle{empty}
\pagenumbering{arabic} 
\setcounter{page}{1}
\renewcommand{\headrulewidth}{1pt}
\fancyhead[LO]{\MakeUppercase\chaptername\ \thechapter}
\fancyhead[RO]{ INTRODUCTION }
\let\Oldpart\part 
\fancyfoot{}
\rfoot{\thepage}

Reviews are statements which express suggestion, opinion or experience
of someone about any market product. On the online e-commerce websites,
users place their reviews on product form to give suggestion or share
experience with product providers / sellers / producers and new purchasers.
The provided user experience can help any business to grow for improvement
by analyzing the suggestions. Polarity of reviews causes certain financial
gain or loss to any product provider. On other side, reviews influence
new purchasers while taking decision of purchasing any particular
product. It can be concluded that effects of reviews target both business
and users in different ways. Keeping this point of view, many firms
/ product providers hire agents to forge fake opinions for growing
their business and market reputation. As a result, users take wrong
product selection decision. 

The pattern of web based shopping is developing day by day. Online
e-commerce websites opened channel for selling or purchasing products.
E-commerce sites facilitates users to purchase product (e.g. motor
bike, headphones, laptop, etc) or avail any service (i.e. hotel reservation,
airline ticket booking, etc). Users often give suggestion/opinion/review/comment
on e-commerce sites to share their experience after using any product
or availing service. Including e-commerce sites, there exist many
blogs created by users which contain user experience of product or
service. The posted reviews/opinions/suggestions provide useful information
to new customers and product providers. Before purchasing any product,
people often use to take suggestion from surrounded people. Those
suggestions help people to decide whether the product is worth purchasing
or not. Our decision of purchasing or not purchasing a product entirely
based on the opinions of people who had used certain product. 

For example, if a person want to buy a smart phone of \textit{samsung},
he would take opinion from his friends or family relatives who had
used smart phone of \textit{samsung}. Purchasing that smart phone
will totally depends on opinions of friends and family relatives.
Opinions or suggestions of people effect decision making of other
people for product selection. Physically, there are limited persons
around us from whom we can take suggestions while choosing a daily
life product. However, it seems very difficult while purchasing a
product if there is no one around us who have used that particular
product. Where as, there are huge number of opinions available on
e-commerce sites and people take their purchasing decision by reading
reviews and product experience of different people. It is nature of
human that more the people favor for purchasing any product more we
can rely on that product. That is why, online reviews can influence
purchasers in a good or bad way. Companies of product hire agents
to distract new customers by placing false opinions about their product
or competent product. The task of identifying fake reviews is very
important for the betterment of new purchasers and good quality product
companies. 

This chapter gives an introduction about our research work. It gives
a brief discussion on trend of purchasing from outdoor to online shopping.
It also discusses about importance of user reviews on a business for
helping new customers and business development. The later part of
this chapter discusses positive and negative effects of fake opinions.
We have also discussed about fake review and their categories. In
addition, an overview of contextual and behavioral features of reviews
and reviewer are discussed. In last, different problems for detecting
fake opinions/reviews and research objectives are defined. 

\section{Importance of User Reviews}

Online purchasers on e-commerce sites are increasing day by day. Online
purchasers often post reviews/opinions about certain product they
have used. In other words, opinions are content created by users on
e-commerce websites to express experience of users about any service
or product. Importance of user reviews can be viewed from user and
business perspective. From user perspective, these reviews can influence
new customers/users for purchasing decision of certain product in
a good or bad way. Decision of new purchasers is influenced by reviews
of users. Good of bad features in accordance with user experience
are described in reviews which help other users for taking the decision
of purchasing the product. For purchasing online, user often visit
e-commerce sites rich with user experience about products. So quality
and number of user experience can effect user traffic on site. 

By looking importance of reviews from business perspective, mining
of user reviews help product providers for improving the business
strategies and product quality. Reviews help product providers for
revealing features demanded by customers which can build the stability
of product in current market. User reviews can impact any company
with certain financial gains or loss. Negative reviews of users can
financially harm the company because new purchasers may divert the
purchasing decision after reading the negative reviews. Positive reviews
on any product influence the purchasing decision of new customers. 

\section{\label{sec:What-is-Fake}What is Fake Review}

Opinion spamming is an immoral activity of posting fake reviews. The
goal of opinion spamming is to misguide the review readers. Users
involved in spamming activity are called ``spammers''. The task
of a spammer is to build fake reputation (either good or bad) of a
business by placing fake reviews. There exist some businesses who
pay spammers to promote the company to attract new customers or to
demote competent company of same type of business. A fake review either
belong to \textit{positive} or \textit{negative} polarity. Review
containing praising statement about the product fall in \textit{``positive
polarity''.} And review containing loathing statements about the
product fall in \textit{``negative polarity''}. 

It is reported by \cite{Sun2016,Jindal2007} after analyzing reviews
on different e-commerce sites that more than ten percent reviews on
e-commerce websites are fake. It was also reported that more than
seventy five percent of fake reviews belong to \textit{positive polarity}.
Increasing need for identifying fake reviews has captured the attention
of researchers for solving the problem. Fake reviews not only mislead
new customer for taking product purchasing decision but also affects
business of good quality product. And due to false and misleading
reviews on particular e-commerce site, users will avoid to visit that
particular e-commerce site. It is concluded that identifying fake
reviews will tackle three loses at one time. 

Fake reviews or spam opinions are classified into three categories
\cite{Jindal2007,Jindal2008}: \textit{untruthful} \textit{reviews},
\textit{brand} \textit{reviews} and \textit{non-reviews}. \textit{Untruthful}
\textit{reviews} are involved in promoting or demoting false reputation
of the particular target product. The content of \textit{untruthful}
reviews may contain statement about different features of product.
\textit{Brand reviews} are posted to assault the product provider,
manufacturer or distributor. Content of \textit{non-reviews} is irrelevant
to product on which the review is posted. Content of \textit{non-reviews}
may contain question-answers or advertisements. 

\section{Fake Review Detection}

It is also called spam opinion detection, fake opinion detection and
spam review detection. The problem of spam opinion detection is classification
problem which separate fake reviews from non-fake reviews. In the
field of machine learning, one of the popular task in supervised learning
is classification. The task of classification is to categorize unknown
objects from pre-defined number of groups/classes/ labels based on
certain properties or features. It is a very difficult to identify
fake reviews by reading huge number of online reviews. With the help
of classification and its variety of techniques we are able to classify
fake reviews from non-fake reviews. The research area of opinion spam
is divided into three tasks: identifying fake reviews, individual
spammers and spammer groups. The focus of our research work is to
identify spam opinions by exploiting different types of features related
to reviewer and review content. Features are characteristics related
to business, reviewer and review. Important information can be extracted
by analyzing these attributes from different perspectives and that
information can reveal the false reviews from huge number of online
reviews. Generally, we classify these extracted features in two categories:
contextual features and behavioral features. 

\section{Contextual and Behavioral Features}

\begin{figure}
\begin{centering}
\includegraphics[scale=0.5]{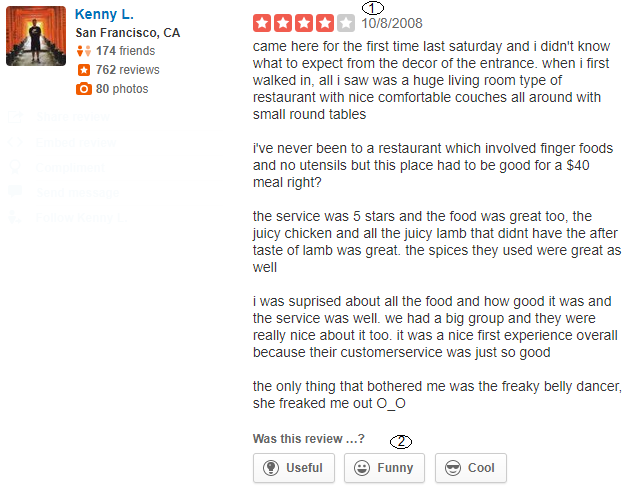}
\par\end{centering}
\caption{\label{fig:Review-On-Amazon}Review On Yelp\protect\textsuperscript{1}}
\textsuperscript{}
\end{figure}

\begin{figure}[tb]
\begin{centering}
\includegraphics[scale=0.5]{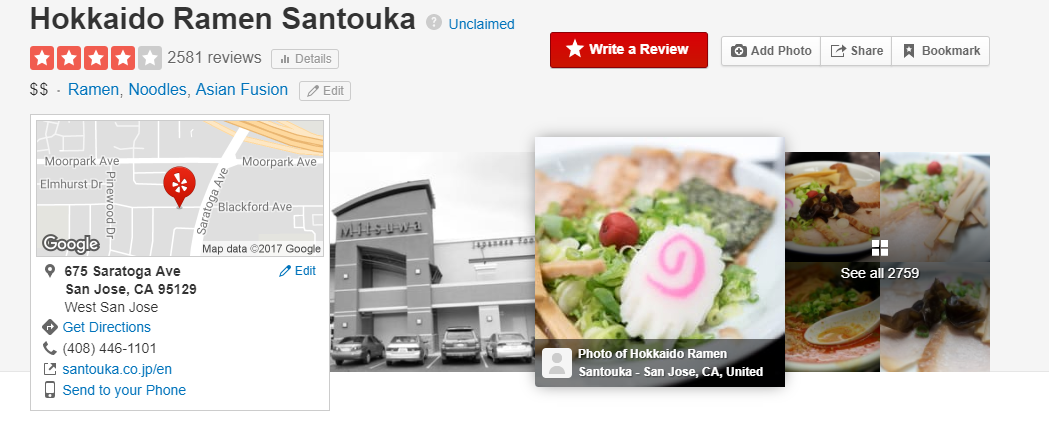}
\par\end{centering}
\centering{}\caption{\label{fig:Product-Information-on}Product Information on Yelp\protect\textsuperscript{1}}
\end{figure}

\begin{figure}
\subfloat[]{\includegraphics[width=2.8in,height=3.5in]{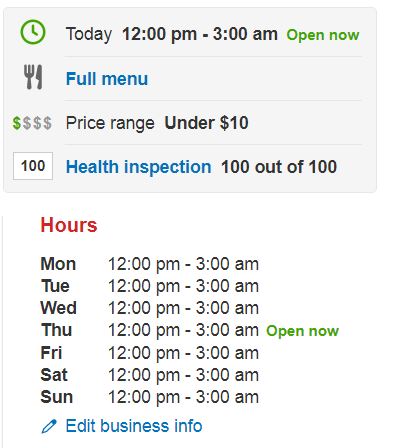}

}\hfill{}\subfloat[]{\includegraphics[width=3in,height=4in]{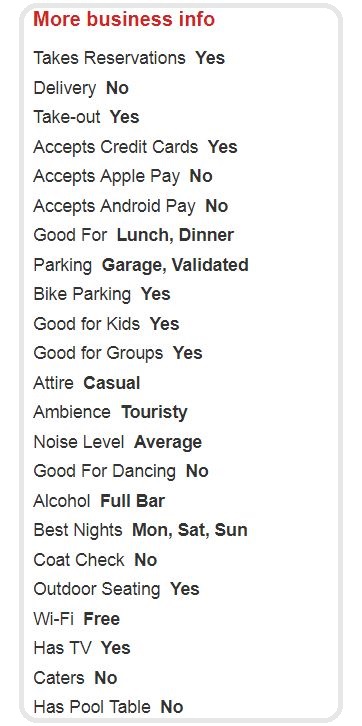}

}
\centering{}\caption{Information about a Restaurant on Yelp\protect\textsuperscript{1}\protect\textsuperscript{}\label{fig:Information-about-a} }
\end{figure}

\begin{figure}
\begin{centering}
\includegraphics[scale=0.5]{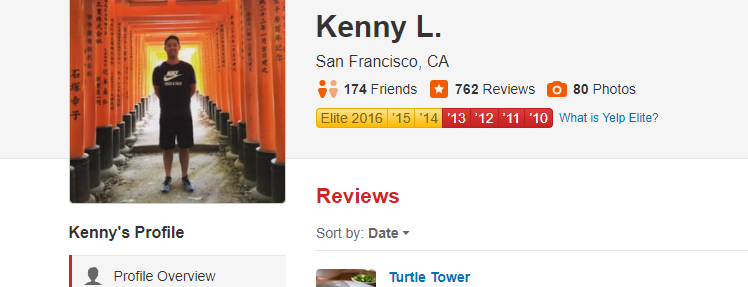}
\par\end{centering}
\caption{\label{fig:User-Profile-on}User Profile on Yelp\protect\textsuperscript{1}}
\end{figure}

It is reported by researchers that the task of identifying untruthful
reviews is more challenging task than identifying brand reviews and
non-reviews \cite{Zhang2016}. Commonly, two types of features are
used to identifying fake reviews: Contextual and Behavioral features.
Both type of features are extracted from three types of attributes:
\textit{review centric attributes, product centric attributes} and
\textit{reviewer centric attributes} \cite{Jindal2007}. A set of
attributes of \textit{review centric attributes} from Yelp\footnote{https://www.yelp.com/sf}
can be seen in Figure \ref{fig:Review-On-Amazon}. Review centric
attributes consist of review content, rating, photos, review post
date, different types of votes and others (Tag 1 and 2 of Figure \ref{fig:Review-On-Amazon}).
Reviewer centric attributes consist of information about reviewer
(e.g. name, location, review count, friend count etc) as shown in
Figure \ref{fig:User-Profile-on}. Product centric attributes contain
information of the product or service (e.g. name, price, brand, description,
etc). Figure \ref{fig:Product-Information-on} and \ref{fig:Information-about-a}
shows product centric attributes. Detailed discussion of all three
types of attributes used in our experimentation is done in Chapter
\ref{chap:Methodology}. 

Contextual features are also called verbal features \cite{Zhang2016}\textcolor{black}{{}
}are extracted from review centric feature. Contextual features represent
different perspective of review content of review. Non-verbal extracted
features are also referred as ``Behavioral features''. Behavioral
features capture unusual behavior of reviewer and content of review
for fake review detection. Extraction of behavioral features is done
from attributes of review, reviewer and product. Exploited contextual
and behavioral features in this research are discussed in detail in
Chapter \ref{chap:Methodology}. Here the main focus of this thesis
is to investigate and exploit those contextual and behavioral features
which improves fake review classification / detection accuracy and
reduce chances of assigning a non-fake review as fake. \blfootnote{https://www.yelp.com/sf}

\section{Motivation and Objectives}

The usage of web platform is expanding and covering every type of
business. The projection of purchasers and sellers is increasing towards
online e-market. The availability of e-market/e-commerce websites
have increased boundary of users for product selection. Whereas, business
of merchants/sellers on e-commerce websites is also growing. It is
natural that people use to ask suggestions from their near ones before
purchasing any daily life product.\textcolor{red}{{} }Physically there
are limited people around whom people can ask but on online platform
there are thousands of reviews/opinion/suggestions of users around
world. The decision of a purchaser totally depends on reviews of users
which effects product provider with certain financial gain or loss.
A review is always true in perception of a purchaser however a common
purchaser does not know that reviews can be faked\textcolor{red}{.}
Users are unbound to give review on any product because there is no
check to see either reviewer has used that particular product or not.
It gives incentives to business to increase their fake reputation
by posting fake reviews to attract new customers. To create fake reputation
for increase product sale in e-market, businesses pay spammers to
post fake reviews. Some businesses pay spammers to harm competent's
reputation in e-market. Many e-market/e-commerce websites are facing
the problem of fake/spam reviews. Harmfulness of fake reviews spread
in three direction including misleading user decision, loss of good
quality product company and due to misleading reviews risk of losing
visitor for e-commerce site. So there is need to identify fake reviews
to help both user and business. Our goal is to identify fake reviews,
however researchers are exploring this area since 2007 but still there
are set of unexplored features which can help to classify fake reviews
from non-fake reviews. Main objective of this report is exploitation
of such features which can bring improvement in accuracy of classification
of fake reviews.\blfootnote{https://www.yelp.com/sf}

\textcolor{black}{The focus of our research work is to identify untruthful
fake reviews. Real life reviews and pseudo fake reviews datasets are
commonly used in experimentation for spam review detection \cite{Heydari2015}.
The real life dataset contains reviews extracted from Amazon, Yelp
etc. The pseudo fake dataset consist of reviews which are created
by researcher through variety of sources. One of the way researchers
has followed to built dataset pseudo fake reviews is annotating fake
and non-fake reviews with the help of annotators\cite{Algur2010}.
Many researchers also adopted methodology of hiring AMT Turkers for
constructing pseudo fake review dataset \cite{Ott2011,MyleOtt,IstiaqAhsan2016,Mukherjee2011}.
AMT is crowd-sourcing platform where businesses (clients) hire turkers
(freelancers) for intelligent tasks. }

It is more difficult to identify fake reviews from real review dataset
than pseudo fake review dataset because real dataset have less accuracy
than pseudo fake reviews. So main objective of this thesis is to exploit
behavioral and contextual features extracted from three type of attributes
to identify fake reviews in real reviews dataset extracted from Yelp. 

\section{Research Questions}

Our focus is to investigate the effect of behavioral and contextual
features of fake review detection. In this thesis mainly following
research questions were thoroughly investigated.
\begin{itemize}
\item What is effect of ``\textit{reviewer deviation}'' with other contextual
and behavioral features to identify fake reviews on Yelp dataset.
\item What is the importance of ``\textit{reviewer deviation}'' compared
with other behavioral features for fake review detection model training
\item Is selected feature set perform well on scaled dataset to identify
fake reviews.
\item What is effect of different weighting schemes calculating the ``\textit{Reviewer
Content Similarity}'' feature of reviewer
\end{itemize}

\section{Research Scope and Limitation}

This work exploited contextual and behavioral of reviews and reviewers
for identifying the fake opinions/reviews. This research work relies
on some suppositions and limitations such as:
\begin{enumerate}
\item This research work is limited to identifying fake reviews written
in English.
\item Research contribution include exploiting contextual and behavioral
features in real life Yelp restaurant and hotel reviews.
\item Some extracted behavioral features relies on attributes available
in Yelp database (e.g. funny count, friend count etc). It is possible
that these attributes may not available on other e-commerce websites.
\item The evaluation of classification model for identifying fake reviews
depends on the labeled database of Yelp reviews. Reviews are labeled
by Yelp spam filtering algorithm.
\end{enumerate}

\chapter{Background Knowledge}

\thispagestyle{empty}
\fancyhead[RO]{BACKGROUD KNOWLEDGE}

Fake review detection task is one of the challenging classification
task in the field of knowledge discovery. Multiple angles of capturing
deception in reviews data have been focused by researchers for a decade.
Focus of our research work is to investigate the techniques and classification
model to identify individual fake reviews by analyzing different perspective
of review data. 

This chapter focuses on background knowledge essential to understand
this research work. Section \ref{sec:Knowledge-Discovery} gives a
brief overview about knowledge discovery. Reviews offer representing
unrevealed hidden knowledge that can be helpful to both consumers
and businesses. Each step in knowledge discovery process has been
discussed. We discuss classification and number of classification
algorithms. In last section, we discuss data distribution problem
for evaluating classification results. 

\section{\label{sec:Knowledge-Discovery}Knowledge Discovery}

\begin{figure}
\centering{}\includegraphics[scale=0.2]{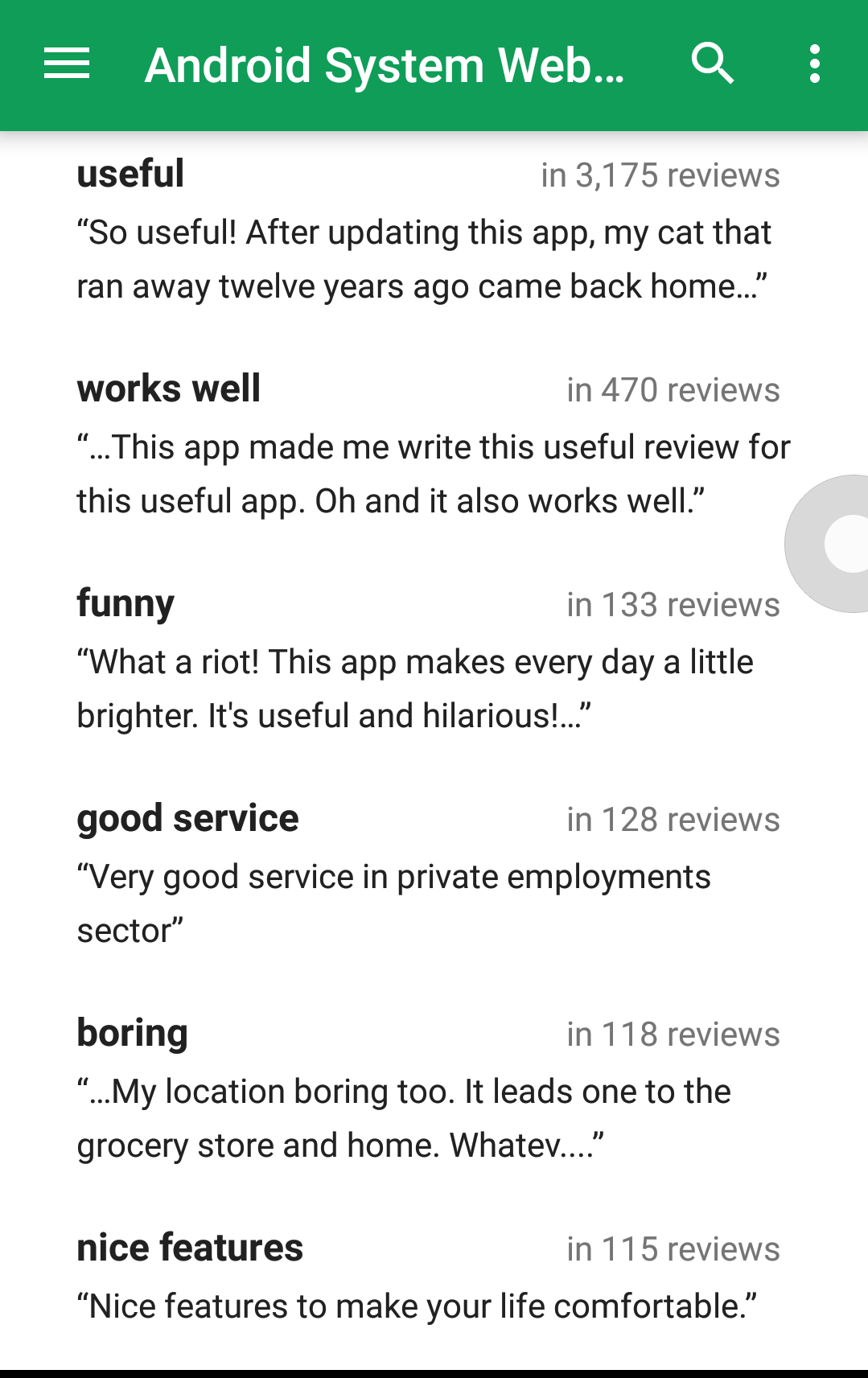}\caption{\label{fig:Classification-of-reviews}Classification of Reviews on
Google PlayStore\protect\textsuperscript{1}}
\end{figure}

\begin{figure}
\begin{centering}
\includegraphics[width=6in]{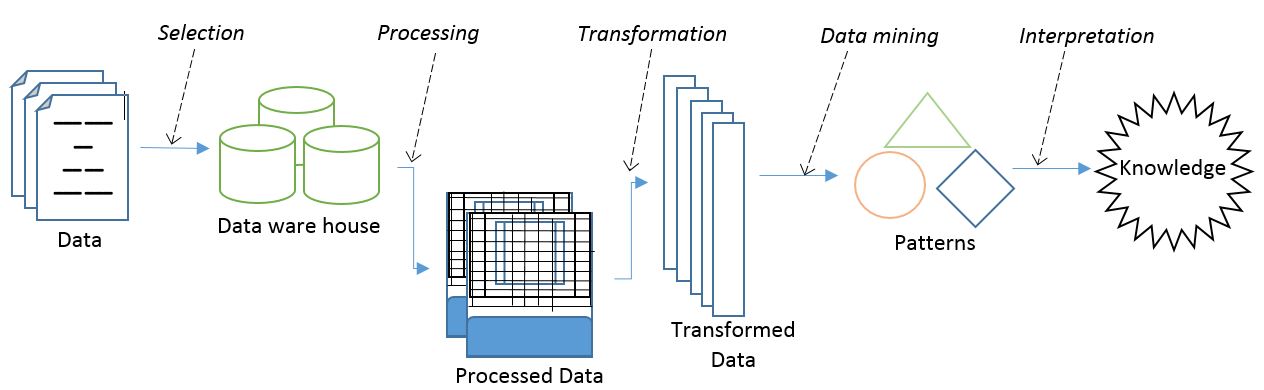}
\par\end{centering}
\caption{\label{fig:Knowledge-Discovery-in}Knowledge Discovery in Database}
\end{figure}

Knowledge discovery is used to extract undiscovered , useful and implicit
information from a large amount of data. The main objective of knowledge
discovery is recognizing patterns in large amount of data. This area
was brought out first by \textit{Frawley et. al }\cite{frawley1992knowledge}
and now it has become one of the most popular research field in computer
science \cite{anand1995evidence}. Access of vast amount of information
for specific application field (like e-commerce, data of genes in
bioinformatics and marketing financial investments) has nourished
the enthusiasm for knowledge discovery greatly. Knowledge discovery
in reviews has also become very important to discover the user interest
and requirement to increase the customer satisfaction and financial
gain. An example of knowledge discovery in reviews is exhibited in
Figure \ref{fig:Classification-of-reviews}. Figure \ref{fig:Classification-of-reviews}
shows that \textit{Goole PlayStore}\footnote{\textit{https://play.google.com/store}}
has categorized the user reviews into different groups by mining the
content and rating of reviews. Iterative process of finding the meaningful
information from raw data is known as process of knowledge discovery.
Stages of the process in database is exhibited in Figure \ref{fig:Knowledge-Discovery-in}.

Data might be gathered from divergent types of sources. There are
commonly two types of data; structured and unstructured. Our selected
dataset contains both structured and unstructured data. The collected
data needs to be cleaned before mining process. Next data cleansing
step requires the existence of noisy, missing and erroneous data values.
It also includes important pre-requisite steps of data reduction like
handling outliers and data type reductions. Subset of data is chosen
based on significance of general information in definite outcome.
Data reduction speed up the data mining algorithms to satisfactory
execution levels especially where large number of attributes are found
in dataset.

After pre-processing next stage is to mine data. Data mining tasks
are based selected discovery goal. Likewise, exploratory examination
the data gives advance bits of knowledge about the nature of data.
Different patterns are explored in data as a result of data mining
process. Recognized patterns are then interpreted and evaluated to
reveal the hidden knowledge of data. This hidden knowledge can be
useful for businesses \cite{anand1995evidence,fayyad1996kdd}. 

\section{Knowledge Discovery Disciplines}

On the basis of diverse formats of data, the research of knowledge
discovery is further divided into different disciplines including
graph mining, image mining, web mining and data mining. \cite{fayyad1996kdd,Heydari2015,Crawford2015}. 

Data represented in the form of graphs are specifically handled by
graph mining. Task of graph mining is to extract patterns of significance
from any graph. Graph mining is introduced by \textit{Yoshida} et.al
\cite{yoshida1994graph}. Transaction graph mining (TGM) and single
graph mining (SGM) are categories of data mining. When you are dealing
with finding patterns in single graph, you have to use SGM however,
TGM is used when set of graphs are dealt\cite{washio2005advances}.
Graph mining has large number of applications in different fields
like web data, biological networks. Graph mining techniques are used
by various researchers to identify the group spammers based on collaboration
of spammer activities. Further usage of graph mining in review data
in discussed in Section \ref{subsec:Identifying-Spammer-Group}.

Image data that is specifically represented in binary form is dealt
by Image mining and also Image mining focuses on image classification,
image mining and image comparison. Satellite, digital images and medical
are some of the examples of binary image data.\cite{wynne2002image}.
It also focuses on image classification, image mining and image comparison.

Recognition of hidden information from web data is focused by Web
Mining. Site page substance, client's information(server logs) and
site page substance are three distinguished types of web data.

Recognition of hidden information from web data is focused in Web
mining. Site page substance, client's information (server logs) and
web hyperlink structures are three types of web data. Three further
subgroups are formed in research of web data mining which are web
content mining, usage mining and web pages structure mining\textbf{.}
Since web pages contain most of the text content so web content mining
is strongly associated with content mining.\cite{cios1998rough}.

Data mining is an important step in the knowledge discovery process
for revealing significant hidden knowledge. This process extracts
the previously unknown information from from huge amount of raw data.
The learning must have the capacity to be utilized by someone having
capability to utilize it , it must be new and non self- evident. Data
Mining (DM) is \textquotedblleft a multidisciplinary field, drawing
work from areas including: database technology, machine learning,
statistics, pattern recognition, neural networks, knowledge-based
system, artificial intelligence, high performance computing, and data
visualization\textquotedblright{} \cite{bramer2007principles}. Data
mining techniques can be applied on a wide range of domains e.g. e-commerce,
business financial studies and bio-informatics due to their adaptive
and versatile nature on different types of data.

\section{Data Mining Techniques}

Generally, DM tasks can be divided into two groups: Descriptive mining
and Predictive mining \cite{fayyad1996kdd,Heydari2015,Crawford2015}.
Descriptive mining involves describing the general characteristics
of the information in the database i.e. clustering and association
rules whereas predictive mining involves forecasting values on the
basis of available current data i.e. regression, classification and
analysis of outlier \cite{berry1997data,han2011data}. We define some
of general techniques of data mining.

\subsection{Clustering}

Clustering is an unsupervised learning technique used to group the
similar type of objects. Unlike classification , we do not need to
train the model in clustering, beside that it's quite similar to classification.
Clustering algorithm is all about discovering the unknown similar
groups. There are many clustering algorithms which are based on calculating
similarity among objects in data collection \cite{han2010fuzzy} e.g.
K-mean clustering, Self Organizing Maps, Fuzzy C-means, Hierarchical
clustering. Clustering is used for many applications in different
domains such as book ordering libraries, city planning, bioinformatics,
image analysis. In the domain of user reviews mining, clustering algorithms
help us to create the clusters of fake and non-fake reviews based
on different identified features of reviews and reviewers. Exploitation
of clustering in identifying deceptive reviews and reviewer by different
researchers is discussed in Section \ref{subsec:Identifying-Spam-Reviews}
and \ref{subsec:Identifying-Spammers}. 

\subsection{Association rules}

An association rule is an if/then statement that defines the uncovered
relationship between seemingly unrelated data in the data collection.
These rules are extracted from the data collection based on how many
times objects appear with each other. The 'if' part contains a single
item whereas 'then' part may contain multiple items. There are two
threshold set on the basis of which association rules are extracted;
one of them is support, support defines the extracted frequency contribution
in the data set whereas the second threshold is Confidence, which
defines the reliability of the extracted rule. There are many association
rules extraction techniques e.g. Apriori algorithm, Eclat algorithm,
FP-growth algorithm. Association rule mining examines the association
analysis of items in market basket.\cite{zaiane1999principles,han2011data}.

\subsection{Outlier analysis}

Another name for outliers is unusual or special cases or astonishment,
they are frequently critical to distinguish. Very often, there exist
data objects that do not comply with general behavior or model of
data. The significance of the outlier analysis depends on domain to
domain. Sometimes outliers are considered as noise in some domains
or it reveals some important information in other one \cite{bramer2007principles}.
Outlier analysis has many applications in different domains, such
as the financial industry, quality control, fault diagnosis, intrusion
detection, Web analytics, and medical diagnosis \cite{aggarwal2017applications}.
Using outlier analysis, various researchers have analyzed behavior
of reviewer to identify spammers \cite{Zhang2006,Jindal2007,Ott2011,Jindal2008,Debarr2009},
because textual or behavioral features of spammers are somehow different
from true ones.

\section{\label{sec:Data-Classification}Classification}

The task of assigning labels to the objects from predefined labels
is known as classification. Classification is a supervised learning
approach that used labeled data to train classification model \cite{bramer2007principles}.Classification
model is able to classify the new objects on the basis of trained
data \cite{fayyad1996kdd,zaiane1999principles,dunham2006data}. There
are three type of classification problem: binomial, multinomial and
multi-label. Binomial classification problem is assigning one label
to unknown object from two classes e.g. Identifying fake and non-fake
reviews. Multinomial classification problem is assigning one label
to unknown object from more than two classes, e.g. reviews are classified
into more than one classes in Figure \ref{fig:Classification-of-reviews}.
Multi-label classification problem is assigning more than one label
to unknown object from multiple classes e.g. scientific document classification.
The fake review detection is classification task to label the reviews
in two classes: \textit{fake and non-fake}. Different approaches are
used by classification algorithm to deal with diverse types of data
collection. Two main areas are included by classification to reveal
the different type of hidden information; text classification and
text mining, which are both central tasks of our research contribution.

\subsection{Data Classification}

Data classification is the field that intersects Information Retrieval
(IR) and Machine Learning (ML) and this field has received a great
enthusiasm in last decade from researchers \cite{hothobrief,osimo2012research}.
. Assigning at-least one precalculated classes of documents having
textual data is the task of data classification. Data classification
is applied in many versatile domains and is an important technique
being used in real life world e.g. Weather forecast, fraud detection,
genes classification and disease diagnosis\cite{dunham2006data,Feldman2007Text,han2011data}.
Apart from these, another important application of data classification
is spam email filtering and review filtering. Different classification
algorithms are proposed to classify object but we are going discuss
some of data classification algorithms related with our research work.

\subsection{Text Classification}

Text Mining(TM) or Text Data Mining is an area of knowledge discovery
which is hot in research as it aims to apply data mining algorithm
to textual datasets. \textquotedblleft It aims at disclosing the concealed
information by means of methods which on the one hand are able to
cope with the large number of words and structures in natural language
and on the other hand allow to handle vagueness, uncertainty and fuzziness\textquotedblright{}
\cite{Feldman2007Text,hothobrief}. Sub-tasks like effort regarding
important portion of text document, text clustering, opinion mining
and document summarization also use text mining for decision making.
\cite{liu2006searching}. Reviews content are in form of text and
extracting different features from reviews fall in the text mining
task. 

\section{Classifiers}

There are many classifiers, some of the include Naive Bayes (NB),
Logistic Regression (LR), K-Nearest Neighbor (KNN), Decision Tree
(DT), Neural Network (NN), Support Vector Machine (SVM), Random Forest
(RF). In this section, classification algorithms are discussed for
building theoretical background so that our proposed approach and
contribution can be understood. Due to best classification results
attained by researchers for out datset, SVM and RF classifiers are
focused in this section \cite{Zhang2016,Mukherjee2013a}. Working
of RF is based on DT due to this reason we are also going to discuss
DT.

\subsection{\label{subsec:Support-Vector-Machine}Support Vector Machine}

In support vector machine (SVM), we arrange related supervised learning
techniques utilized for regression and classification. Given a set
of training examples, we can say that each set is distant as having
place with one of two classifications, a model is assembled by the
SVM preparing calculation which predicts if another object falls into
same classification or the other. Intuitively, with the motive to
find possible wide space between cluster of points, SVM places object
in the space as a point. Mapping of new objects is placed on same
space. Predicted objects to have a place with a classification in
view of which side of the separated space they fall on \cite{duda1973pattern,witten2016data}.
In SVM, dimensions in the space refer to the attributes of the object.
SVM is used to create a hyper-plane in the multidimensional space
exhibiting all the attributes of the object. Instinctively, The more
is the space between two classes , the lesser chances in generalization
error of the classifier. 

\begin{figure}
\begin{centering}
\subfloat[Linear]{\includegraphics[scale=0.6]{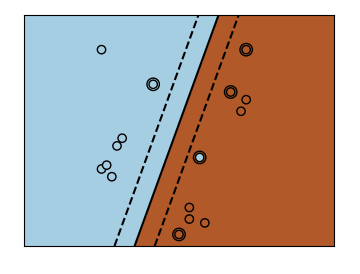}

}\subfloat[RBF]{\includegraphics[scale=0.6]{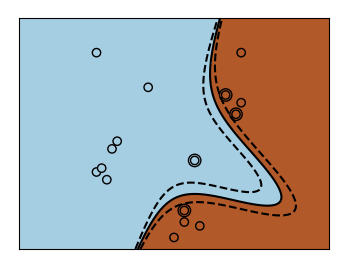}

}\subfloat[Polynomial]{\includegraphics[scale=0.6]{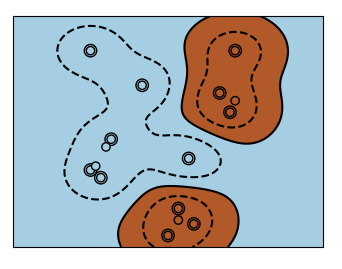}

}
\par\end{centering}
\caption{SVM Kernels\protect\textsuperscript{1} \label{fig:SVM-Kernels}}
\end{figure}

\begin{figure}
\begin{centering}
\includegraphics[scale=0.7]{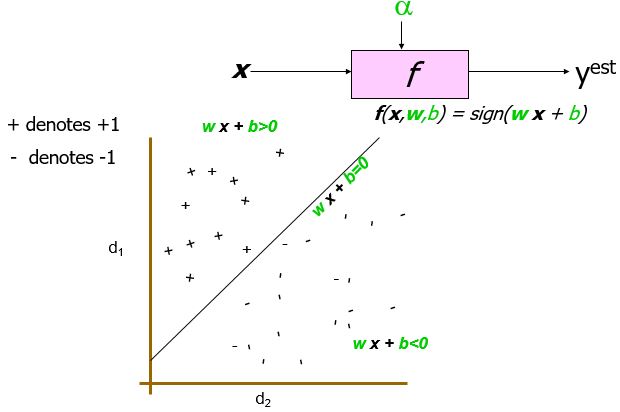}
\par\end{centering}
\caption{Linear Kernel Example\protect\textsuperscript{1} \label{fig:Linear-Kernel-Example}}
\end{figure}

To specify the pattern of constructing space between classed, three
kernel functions are used, Linear, RBF and Polynomial as exhibited
in Figure \ref{fig:SVM-Kernels}\footnote{http://scikit-learn.org/stable/auto\_examples/svm/plot\_svm\_kernels.html}.
Sometimes, object-points are not linearly seperable, in these cases,
the RBF and polynomial are particularly used. Linear SVM uses the
Equation \ref{eq:svm}, where vector of feature of document is represented
by 'x' , weight of 'x' is represted by 'w' , and tunning parameter
represented by 'b'. An example of linear kernel SVM is shown in Figure
\ref{fig:Linear-Kernel-Example}, there are two dimensional data objects.
Two type of classes on space are separated by hyper-plane. \blfootnote{http://scikit-learn.org/stable/autoexamples/svm/plot-svm-kernels.html}

\begin{equation}
y=w.X-b\label{eq:svm}
\end{equation}

The classification of SVM is effective on high dimensional spaces.
Also, SVM performs well on the clear margin of separation. Classification
is also effective when number of dimensions is greater than number
of samples. However, the performance of SVM is not good when data
is noisy. Many researchers have also used SVM in text classification
for fake review detection Text classification tasks having unbalanced
training examples have also used SVM classifier. Hence, we have chosen
this to be used in our research experiments.

\subsection{Decision Trees}

At the start when artificial intelligence field was emerging, a large
set of rules were used to process a large scale dataset on the machine.
As the data set size grew larger, Designing the number of rules in
accordance with data became difficult . This is when decision tree
were invented to solve the problem by creating a tree of rules from
which new instances can be assigned predefined class \cite{Safavian1991}.
To improve the efficiency and to make the diverse data more accurate,
decision trees are designed. \cite{dattatreya1985decision,chang1977fuzzy}. 

Working of a decision tree goes like this; each node is the data feature
whereas conditions are represented by diamond (i.e. value greater
than , less than or equals to 6) over feature of node. New instances
are assigned to the ending node known as leaves. Decision trees are
used in a variety of domains to classify data and to create rules
for large data set and complex feature rules are broken down into
simple ones For example, disease diagnosis, speech recognition, expert
systems, radar signal classification and many more

Order of the nodes is assigned according to the importance score of
the data features while designing the tree. Multiple algorithms is
proposed to split a node for constructing decision tree. Decision
tree splits the nodes on all available attributes and then selects
the split which results in most homogeneous sub-nodes. Algorithm for
construction tree is also based on type of data. Here, we are going
to discuss only 'Gini Index'. This algorithm is used inside selected
library in experimentation.

\subsubsection*{Gini Index}

It says that if we randomly select two items from a population then
both items must be of same class and if population is pure then probability
for this case is 1. There are two steps to calculate \textit{Gini
\cite{fayyad1992attribute}} for split:
\begin{enumerate}
\item Calculate \textit{Gini} for sub-nodes using Equation \ref{eq:gini}.
\item Calculate \textit{Gini} for split using weighted \textit{Gini} score
of each node of that split. 
\begin{equation}
w=p^{2}+q^{2}\label{eq:gini}
\end{equation}
\end{enumerate}

\subsection{\label{subsec:Random-Forest}Random Forest}

Random forest is the collection of decision trees or we can say it
makes the forest of decision trees. It can be utilized for both classification
and regression. To achieve high robustness and accuracy, more trees
in the forests are required. Random forests are constructed using
the same method of constructing decision trees. Multiple trees are
constructed independently and parallel. All the training instances
examined with substitution are utilized while constructing each tree.parameters
at every node of tree is enhanced by constructing the forest of decision
trees.

\begin{figure}
\begin{centering}
\includegraphics[scale=0.35]{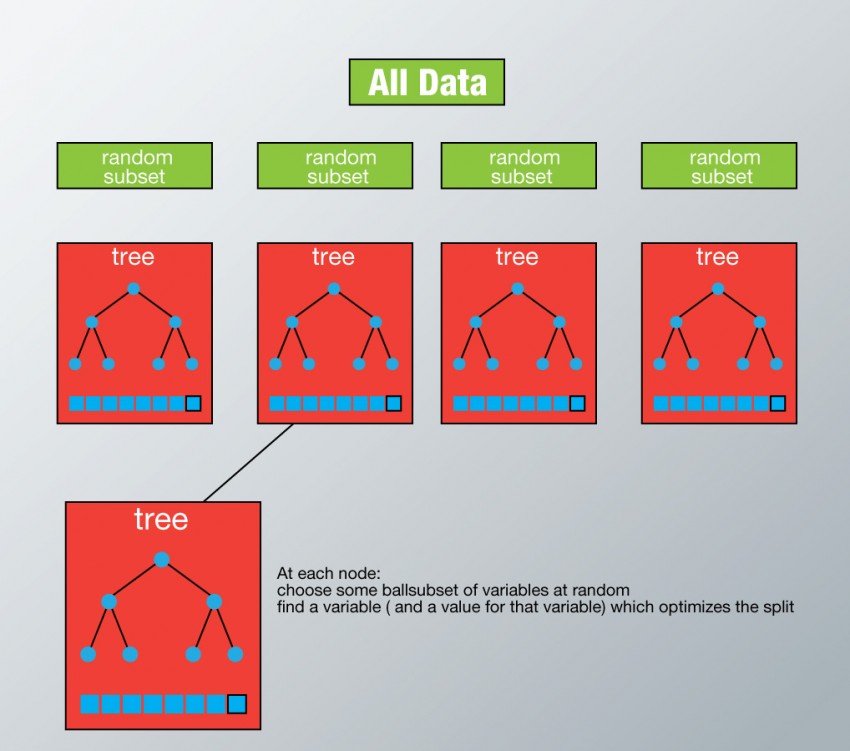}\caption{Working of Random Forest\protect\textsuperscript{3} \label{fig:Working-of-Random}}
\par\end{centering}
\end{figure}

To enhance the parameters at every node of tree, the forest of decision
trees is constructed. A random subset of the set of features is approached
by every node of the tree while training on independent tree as exhibited
in the Figure \ref{fig:Working-of-Random}\footnote{https://www.analyticsvidhya.com/blog/2016/04/complete-tutorial-tree-based-modeling-scratch-in-python/}.
Only one randomly chosen subset of the entire set of features is accessible
to each node of the tree. \cite{schapire1998boosting,liaw2002classification}. 

It can handle large dataset with high dimensional. It also perform
well if there are missing values in dataset. It can be used for outlier
detection and extended for unsupervised learning. We can apply random
forest to classification as well as regression as it performs well
at highly dimensional data set. The usefulness of Random forests allow
them to apply it in various domains like for example banking domain
for fraud detection, stock exchange market to analyze behavior, disease
diagnosis,medicine component composition,  recommendation in e-commerce
etc. Having versatile nature and pros, there are still some cons of
random forest. It does not perform well for regression compared with
classification as it does not give precise continuous nature predictions.
Random Forest behave like a black box approach for statistical modelers.
We have limited control on what the model does.

\section{\label{sec:Classifier-Evaluation}Classifier Evaluation}

\blfootnote{https://www.analyticsvidhya.com/blog/2016/04/complete-tutorial-tree-based-modeling-scratch-in-python}Measures
for Classification Performance is basic part in directing the training
classifiers. Assessment techniques and measures are important as classification
algorithm and are the main key stage to an effective data mining.
There are various criteria for evaluating classifiers and criteria
is set based on selected goal. This research focuses on accuracy of
classification algorithms. Many evaluation measures are used in fake
review detection research area. We define some evaluation measure
used in this research include $Precision$, $Recall$, $F1-measure$
and $Accuracy$ \cite{banker1989sensitivity}. In literature, these
four measures are commonly used for assessing classification model
\cite{Zhang2016,Mukherjee2013} (defined in Equations \ref{eq:precision}-\ref{eq:accuracy}).
We have evaluated our trained classifiers with defined four evaluation
measures. Number of fake and non-fake reviews correctly classified
are denoted by $t_{fake}$ and $t_{non-fake}$ respectively. Whereas,
incorrectly classified fake and non-fake reviews are denoted by $f_{fake}$
and $f_{non-fake}$. Values of these evaluation measures range from
0 to 100. 

\begin{equation}
Precision=\frac{t_{fake}}{t_{fake}+f_{fake}}\label{eq:precision}
\end{equation}

\begin{equation}
Recall=\frac{t_{fake}}{t_{fake}+f_{non-fake}}\label{eq:recall}
\end{equation}

\begin{equation}
F1=\frac{2(Precision.Recall)}{Precion+Recall}\label{eq:f1}
\end{equation}

\begin{equation}
Accuracy=\frac{t_{fake}+t_{non-fake}}{t_{fake}+t_{non-fake}+f_{fake}+f_{non-fake}}\label{eq:accuracy}
\end{equation}

\section{\label{sec:Imbalance-Class-Distribution}Imbalance Class Distribution
Problem and Its Solution}

\begin{figure}
\begin{centering}
\includegraphics[scale=0.7]{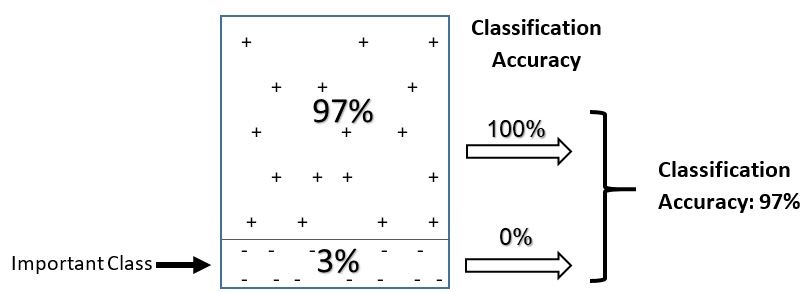}
\par\end{centering}
\caption{\label{fig:Imbalanced-Class-Distribution}Imbalanced Class Distribution
Problem}
\end{figure}

Commonly, the performance of classification algorithms is well when
the training data contain equal number of instance of each class in
dataset. But many real life dataset contain the unequal number of
instance in classes. The outcomes of classifiers may be inaccurate
in certain situation \cite{phung2009learning}. This problem of data
distribution is know as ``\textit{Imbalance Class Distribution Problem}''
\cite{chen2008information,he2009learning}. This problem happens when
a class (\textit{majority class}) contains very high number of instances
compared with another class (\textit{minority class}) in dataset.
A classifier typically has a tendency to decide on majority class
and ignore the minority class for this situation . Figure \ref{fig:Imbalanced-Class-Distribution}
illustrates this problem where 97\% of instances relates to majority
class and 3\% instances of minority class. 

The issue of imbalance class distribution is unavoidable. In current
days, many domains are facing class imbalanced data nature. Imbalance
class distribution create hurdles to classifiers in learning. In our
domain of fake review detection, the number of true reviews are immensely
high than fake reviews as reported by\cite{Jindal2007,Sun2016} that
e-commerce sites contains around 10\% of fake reviews. One of the
popular technique to solve this problem is ``\textit{sample technique}''
which is defined below.

\subsection{Sampling Techniques}

Sampling technique is a popular approach to tackle imbalance class
distribution problem. Distribution of minority and majority class
are altered for training. It follows equal number of instances in
both majority and minority class \cite{phung2009learning}. There
are two techniques to overcome imbalance class distribution: \textit{under-sampling}
and \textit{over-sampling}.

\begin{figure}[t]
\begin{centering}
\includegraphics[scale=0.8]{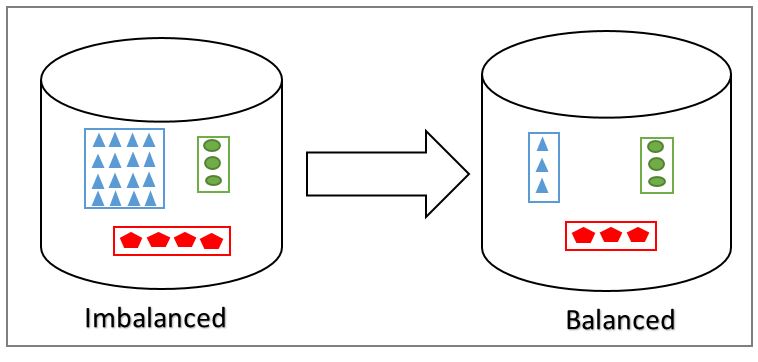}
\par\end{centering}
\caption{\label{fig:Under-sampling-approach}Under-sampling Approach}
\end{figure}

\begin{figure}
\begin{centering}
\includegraphics[scale=0.8]{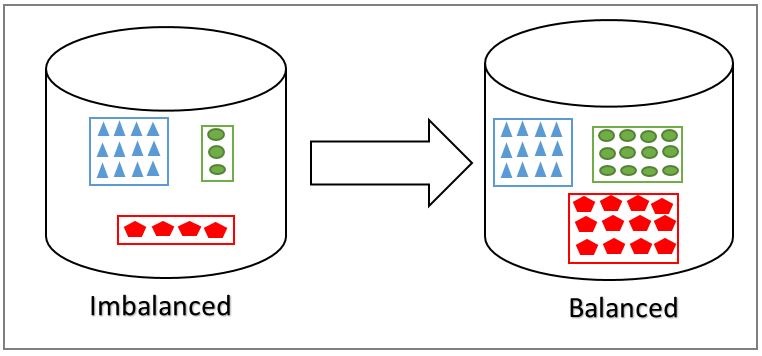}
\par\end{centering}
\caption{\label{fig:Over-sampling-approach}Over-sampling Approach}
\end{figure}

\subsubsection{Under-Sampling}

This technique decrease the number of majority class samples. Those
samples are selected randomly \cite{he2009learning}. The aim of under-sampling
approach is to reduce the skewed distribution of minority and majority
class by downsizing of majority class \cite{yen2009cluster}. This
approach is appropriate in huge dataset where instances of minority
class is very less than majority class \cite{hu2009msmote}. Process
of under-sampling approach is illustrated in Figure \ref{fig:Under-sampling-approach}.
The figure shows three types of classes occur in dataset and after
under-sampling each class contain equal size of instances for training.
In our domain of reviews we follow this approach to select randomly
same number of reviews from both fake and non-fake labeled reviews.

\subsubsection{Over-Sampling}

Unlike under-sampling, the examples of minority class are increased
in this approach. Increasing the samples include replication of random
examples of minority class. Applying under-sampling posits the chance
of losing information about data but this approach would not compromise
information losing. However, computational cost is increased in this
technique. Figure \ref{fig:Over-sampling-approach} shows working
of this technique, where size of minority class is increased by replicating
the training examples. 

\chapter{\label{chap:Related-Work}Related Work}

\thispagestyle{empty}
\fancyhead[RO]{RELATED WORK }

The research area of opinion spam detection is a challenging task
since ten years. First investigation on analyzing spamming activities
in reviews was studied by Jindal \cite{Jindal2007}. Three research
areas in fake/spam review detection were discussed which include:
identification of fake reviews, Individual Spammer and Spammer Group.

First and most common investigation research area is to identify individual
fake reviews. Second type of investigation reveals the user accounts
that are involved in the deception activity of posting fake reviews
called \textit{``spammer detection}''. Third type of investigation
identify the groups of users that involved in having activity of posting
fake reviews to achieve single goal called \textit{``Spammer Group
Detection}''. Sixty two percent of publications have focused on detection
fake reviews. Thirty one and 7 percentage of research work was on
individual spammer and spammer group detection, respectively. Our
research work targeted on identifying fake reviews using classification
method. Mainly two types of datasets are used for experimentation
in identifying fake reviews: real life and pseudo fake as shown in
Table \ref{tab:Research-Work-Reported}. Two types of features are
reported in classifying fake reviews: contextual and behavioral. Our
research work focuses on identifying spam reviews using real life
dataset on contextual and behavioral features. This chapter gives
a brief overview of research work carried out in identifying spam
reviews, individual spammers and group spammer.

\section{\label{subsec:Identifying-Spam-Reviews}Identifying Spam Reviews}

A quick overview of literature related to identifying fake reviews
can be seen in Table \ref{tab:Research-Work-Reported}. \textit{Jindal}
et.al reported a preliminary investigation on spam opnion detection
\cite{Jindal2007}. In their next publications, detailed analysis
of fake reviews on Amazon were reported \cite{Jindal2008,Jindal2007a}.
They considered 5.8 million reviews from Amazon \footnote{https://www.amazon.com/}
and used feature from product and reviewer meta-data on four categories
of product to identify fake reviews. The research work comprises identification
of untruthful reviews, brand reviews, non-reviews and spammer groups.
They discovered spamming activities including identifying duplicate
or near duplicate reviews using shingle method. Untruthful reviews
were identified by calculating content similarity between all reviews
of a reviewer to highlight duplicate reviews. For identifying brand
reviews and non-reviews, dissimilarity between product meta data and
review content were used. Spammer groups were identified by calculating
content similarity of reviews of different reviewers. Logistic Regression
was adopted for training on duplicate reviews which achieved 78\%
on AUC (Area Under ROC curve).

\begin{table}[h]
\begin{centering}
\begin{tabular}{|c|>{\centering}p{0.9in}|>{\centering}p{1in}|>{\centering}p{1in}|>{\centering}p{0.8in}|>{\centering}p{0.8in}|}
\hline 
\textbf{Year} & \textbf{Author} & \textbf{Dataset Type/Source} & \textbf{Classifier} & \textbf{Feature Type} & \textbf{Evaluation Measures}\tabularnewline
\hline 
\hline 
2007 & Nitin Jindal \textit{et. al} & Pseudo Fake/Amazon & LR & Contextual & AUC\tabularnewline
\hline 
2008 & Nitin Jindal \textit{et. al} & Pseudo Fake/Amazon & LR & Contextual & AUC\tabularnewline
\hline 
2010 & C. Lai \textit{et. al} & Pseudo Fake/Amazon & SVM & Contextual & AUC, Recall Precision\tabularnewline
\hline 
2010 & Siddu Algur\textit{ et. al} & Pseudo Fake/ Web Page & - & Contextual & Accuracy\tabularnewline
\hline 
2011 & Fangtao Li\textit{ et. al} & Pseudo Fake/Epinions & LR, SVM, NB & Contextual & Precision, Recall, F1\tabularnewline
\hline 
2013 & Mukherjee Arjun\textit{ et. al} & Real life/Yelp & SVM & Contextual, Behavioral & Precision, Recall, F1, Accuracy\tabularnewline
\hline 
2014 & H. Li \textit{et. al} & Real life/Diaping & SVM & Contextual, Behavioral & Precision, Recall, F1\tabularnewline
\hline 
2014 & Yuming Lin \textit{et. al} & Pseudo Fake/Amazon & LR, SVM & Contextual, Behavioral & Precision, Recall, F1\tabularnewline
\hline 
2016 & Istiaq Ahsan \textit{et. al} & Pseudo Fake, Real Life/ AMT+Yelp & NB, SVM & Contextual & Precision, Recall, F1, Accuracy\tabularnewline
\hline 
2016 & Dongsong Zhang \textit{et. al} & Real life/Yelp & SVM, DT, RF, NB & Contextual, Behavioral & Precision, Recall, F1, Accuracy\tabularnewline
\hline 
\end{tabular}
\par\end{centering}
\caption{\label{tab:Research-Work-Reported}Research Work Reported in Identifying
Fake Reviews}
\end{table}

In \cite{Algur2010}, two annotators were hired to construct pseudo
fake reviews dataset containing 960 reviews. The proposed research
work comprises identifying duplicate and near duplicate reviews using
humming distance. Fifty seven percent accuracy was reported in proposed
technique. However, reported accuracy was not significant but their
feature extracted from reviews was novel. 

Reviews from Epinions\footnote{http://www.epinions.com/} was annotated
to build dataset by \cite{li2011learning}. Several features similar
to \cite{Jindal2008} are proposed with other features including authority
score calculated using PageRank, positive and negative polarity of
reviews. They trained SVM, LR and NB to classify fake reviews. Reported
results showed that NB achieved best F-Score (58.30\%). 

\cite{Ott2011} hired AMT \footnote{https://www.mturk.com/mturk/welcome}
turkers to create dataset of hotel reviews. Annotated dataset of reviews
contained 400 fake and 400 non-fake reviews. SVM and NB were trained
on three types of features. First, Parts of Speech (POS) tagger, n-gram,
and features from Linguistic Inquiry and Word Count (LIWC) were used.
Four approaches were adopted to classify spam reviews. SVM trained
on bigrams and LIWC extracted features achieved 89.8\% accuracy. 

\cite{Wu2010} forged fake reviews for TripAdvisor\footnote{https://www.tripadvisor.com/}.
An unsupervised learning approach based on ranking of product was
proposed to identify spam reviews. Fake reviews may damage product
ranking on TripAdvisor. It was reported that proposed approach based
on product ranking is effective for identifying fake reviews.

\cite{Lai2010} proposed feature set to identify untruthful and non-reviews
category of fake reviews. Feature set for identifying non-reviews
includes lexical, syntactical and stylistic features. For identifying
untruthful reviews, dataset was build containing reviews from Amazon.
Two annotators were hired to annotate subset of crawled reviews. SVM
acquired 96\% recall in classifying non-reviews. Reported results
showed improvement in AUC including recall. Using three types of contextual
features with KL-divergence, untruthful reviews were identified on
hyper-plane. 

\cite{Li2014a} used dataset of chinese reviews from ``Diaping''.
On basis of behavioral feature ``Diaping'' is filtering fake reviews.
SVM and Positive-Unlabeled (PU) learning are used to improve classification
model. Positive refers to fake reviews and unlabeled refers to unclassified
fake and non-fake reviews. In PU learning, model can be trained only
on fake reviews. The classification results reported improvement in
recall value upto 89\%.

Based on three features of review text similarity and two features
of reviewer's posting rate \cite{LinY2014} proposed accumulative
formula to separate fake and non-fake reviews. To calculate the similarity
between all reviews, reviewer's reviews and reviews of a product Jaccard-Similarity
was computed. Reported results shows that SVM outperformed LR on dataset
adopted by \cite{Jindal2008,Jindal2007}. The trained SVM classification
model achieved 85\% F1-score.

Real life review dataset or pseudo fake review dataset are used for
experimentation. \cite{IstiaqAhsan2016} construct dataset by combining
real life reviews and pseudo fake reviews. Unlabeled dataset contains
reviews from Yelp and Labeled dataset of \cite{Ott2013} were used.
Significant results were reported by using hybrid dataset. The basic
novelty of proposed word was applying active learning with supervised
learning. Duplicate reviews were identified from unlabeled dataset
using KL-JS distance. SVM was trained on duplicate reviews to label
reviews of real life dataset. SVM placed reviews on hyper-plan to
separate reviews. Manual classification of reviews (close to hyper-plane)
by user was done. The accuracy of 88\% was reported using NB. 

\cite{Mukherjee2013a} exploited contextual and behavioral feature
to train classification model. Four unexploited behavioral features
were explored for fake review detection. It was empirically proved
that using only contextual features can obtain 68.1\% accuracy for
pseudo fake reviews. Classification model trained on contextual features
do not achieve significant accuracy for real life fake reviews. \cite{Mukherjee2013a}
justified that AMT Turkers are not good at faking a review. The reason
is that AMT Turkers have limited knowledge about the domain. Word
distribution of posted reviews by AMT turkers\textquoteright{} is
different from true reviewer. This was the reason \cite{Ott2011}
reported high accuracy of classification model trained on contextual
features. \cite{Mukherjee2013a} trained classification model on real
life reviews (from Yelp) using combined contextual and behavioral
features. The selected feature set consist of ``\textit{reviewer
deviation}'', ``\textit{positive ratio}'', ``\textit{maximum posting
rate}'', ``\textit{review length}'', ``\textit{average content
similarity}'' and n-grams features. Reported results showed 86\%
accuracy on restaurant reviews. It was also reported that 75\% of
spammers post atleast 6 reviews in a day. Their analysis reported
that 80\% of spammer are hired for promoting the reputation of business.
Third finding of the analysis reported that 70 out of 100 spammers
have text similarity between their posted reviews.

Various behavioral features related with reviews and reviewer were
exploited by \cite{Zhang2016} . The importance of selected features
were also investigated for identification of fake reviews. They exploited
features including 24 behavioral and 16 contextual features. Exploration
of behavioral features were based on Interpersonal Deception Theory
(IDT) which posits that \textquotedblleft deceivers display both strategic
behaviors (e.g., information manipulation) and nonstrategic behaviors
during deception\textquotedblright{} \cite{buller1996interpersonal}.
Experiments were conducted on review dataset of Yelp. Reported classification
results show 87.8\% accuracy using all features. Highly correlated
features were identified using Pearson Correlation before feature
pruning. Twelve most important features were identified after feature
pruning to achieve 90\% accuracy by training RF classifier. Moreover,
accuracy of SVM, NB, Decision Tree and RF were also compared. 

We have discussed both supervised and unsupervised learning approaches
used to identify the fake reviews. To the best of our knowledge, supervised
learning approaches acquired better results than unsupervised. Therefore,
supervised learning is dominant over unsupervised in this research
area. It can also analyzed from literature that limited research work
is done on real life review dataset. And combining contextual and
behavioral features can improve classification model for identifying
fake reviews. \cite{Mukherjee2013a} and \cite{Zhang2016} exploited
both contextual and behavioral features for classifying fake and non-fake
reviews. However, selected feature set by \cite{Zhang2016} is different
from \cite{Mukherjee2013a}. Here, we find that combining both feature
set (with feature pruning) is worth investigating to improve the classification
model.

\section{\label{subsec:Identifying-Spammers}Identifying Individual Spammer}

A graph based approach was proposed by \cite{Wang2011}. Reviewers,
reviews, and products were considered as nodes. Based on various features,
the edges between reviewer, review and product were placed. Proposed
technique captured relationship between nodes ``\textit{trustworthiness}'',
``\textit{product reliability}'' and ``\textit{honest polarity}''.
The ``\textit{trustworthiness}'' of reviewer is based on ``\textit{honest
polarity}'' of posted reviews. The ``\textit{honest polarity}''
is accumulative score of ``\textit{product reliability}'' and ``\textit{review
deviation}'' within a specific time frame. A product is considered
as reliable if certain number of posted reviews (positive) belong
to trustworthy reviewers. Based on these scores, an iterative algorithm
was proposed to assign score to reviewer in range from 1 to -1.

Many researchers assumed that spammers usually allocate a particular
time interval to place fake reviews. This assumption was used to identify
spammers account \cite{xie2012review}. The capturing of unusual events
and numbers of reviews rise dramatically in that interval. The behavior
of the reviewer was analyzed in spam attacks, casual purchasing and
promotion. It was analyzed that spammers start posting reviews as
soon as they are hired. Dividing the reviewing duration into time
frames may help to detect these type of spam attacks. Three algorithms
were used which include: i) Bayes change point detection algorithm
ii) template matching algorithm for finding burst patterns and iii)
a sliding window to detect blocks in time series matched with a joint
burst in all dimensions of the time series. 

The review burst is also focused by \cite{Fei2013} which included
other behaviors of the reviewer. Reviews from Amazon were considered
for the study that included beahvior of reviewer like rating deviation,
ratio of Amazon verified purchasers, content similarity between reviews
of a reviewer and burst review ratio. The Loopy Belief Propagation
Algorithm Loopy was used to process these features in the network
of reviewers. The overall 57.5\% accuracy was reported with the followed
experimentation.

\cite{Akoglu2013} proposed framework ``\noun{FraudEagle}'' based
on graph to identify spammers and false ratings. Unsupervised learning
approach was adopted in proposed framework. Signed bipartite graphs
were used to classify the network of spammer and non-spammer. The
proposed framework ``\noun{FraudEagle}'' focused on connectivity
between review content. Proposed framework also analyzed sentiment
orientation of review text on a product. 

\section{\label{subsec:Identifying-Spammer-Group}Identifying Spammer Group}

So far research on two areas of fake review detection has been discussed
in previous sections (Section \ref{subsec:Identifying-Spam-Reviews}
and Section \ref{subsec:Identifying-Spammers}). However, identifying
spammer group is different task. Here, the focus will be on research
reported in identifying spammer group. 

Different firms may utilize various spammers in a group to promote
or demote a product. An interesting research in this regard is carried
out by \cite{liu2012sentiment}. It is discussed that the activity
of group spamming can be carried out in two ways. First, an isolated
spammer use more than one account to post fake reviews. Second, more
than one spammers are hired by a business to promote or demote any
target item \cite{zhang2015detecting,ye2015discovering}.

Many spammers might be hired independently for assaulting any target
product. By following this manner, capacities of spamming can be expanded.
First labeled dataset for spammer group detection was constructed
by \cite{Mukherjee2012}. The dataset contained labeled 2412 non-spammer
and spammer groups. Feature used by \cite{Mukherjee2011} were improved
to identify spammer groups. The selected Feature set consist of ``\textit{group
content similarity}'', ``\textit{reviewer burstiness}'', ``\textit{group
deviation}'', ``\textit{average content similarity}'' and others.
Spam score of identified groups is calculated by combing all features
using frequent-pattern mining. Relational model based GRank was used
to to rank all groups to identify spammer groups.

\chapter{\label{chap:Methodology}Research Methodology}

\thispagestyle{empty}
\fancyhead[RO]{RESEARCH METHODOLOGY }

\begin{figure}
\begin{centering}
\includegraphics[scale=0.5]{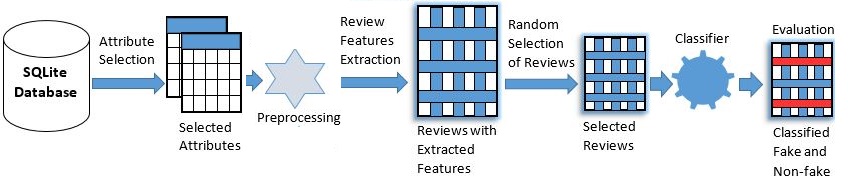}
\par\end{centering}
\caption{Steps in the Process of Fake Review Detection \label{fig:Steps-in-the}}
\end{figure}

This chapter discusses our research methodology. Figure \ref{fig:Steps-in-the}
gives a brief overview of our research methodology. This chapter focuses
on source and attributes of dataset used in this research. It also
focuses on contextual and behavioral features. In later part of this
chapter, some classification model and evaluation measures is briefly
discussed. Our research methodology includes following steps: 
\begin{flushleft}
\textbf{Dataset selection:} Different domains of review data are analyzed.
Selection of data include restaurant and hotel reviews from Yelp\footnote{https://www.yelp.com/sf}. 
\par\end{flushleft}

\begin{flushleft}
\textbf{Preprocessing: }Preprocessing techniques are used to handle
noisy and inconsistent data. Main preprocessing techniques that were
applied include tokenization, lemmatization and others.
\par\end{flushleft}

\begin{flushleft}
\textbf{Feature Extraction:} After preprocessing contextual and behavioral
features are extracted from available attributes in review database
to make feature set for classification model.
\par\end{flushleft}

\begin{flushleft}
\textbf{Model Training:} Different classification models are then
trained for experimentation related to our research. Main focus was
on SVM and RF.
\par\end{flushleft}

\begin{flushleft}
\textbf{Evaluation and Analysis:} All outcomes from different classification
model are then evaluated using different evaluation measures.
\par\end{flushleft}

Supervised and unsupervised learning approaches for identification
of fake reviews are discussed in related work (Chapter \ref{chap:Related-Work})
. To the best of our knowledge, supervised learning approaches acquired
better results than unsupervised learning \cite{Heydari2015,Mukherjee2013}.
Therefore, supervised learning is dominant over unsupervised that's
supervised learning is used. For fake review detection, proposed research
work include analyzing contextual and behavioral features of reviews
and reviewers. We used reviews data from Yelp for experimentation.
Contextual and behavioral features are extracted from review dataset. 

Twelve features (including contextual and behavioral) were selected
by\cite{Zhang2016}. More than 30 features including contextual and
behavioral features were explored on real life dataset \cite{Zhang2016}
but selected twelve features obtain high importance score and classification
accuracy. We combine \textit{``Reviewer Deviation}'' explored by
\cite{Mukherjee2013a} with feature set of \cite{Zhang2016} for training
classification model to identify fake reviews. To the best of our
knowledge, selected feature set is not exploited in literature for
fake review detection. Therefore, the effect of this feature is explored
with other behavioral and contextual features. 

Entities and attributes in real life dataset considered for experimentation
in Section \ref{sec:Dataset} in this chapter. Section \ref{sec:Preprocessing}
discussed preprocessing techniques used for experimentation purpose.
Different types of features extracted from available attributes are
discussed in Section \ref{sec:Features}. Classifiers and some evaluation
measures used in this research are briefly discussed in Section \ref{sec:Classifiers-Adopted}
and Section \ref{sec:Evaluation-Measures}.

\section{\label{sec:Dataset}Dataset}

Most e-commerce sites (e.g. Yelp.com, Epinions.com and Amazon.com)
allow users to place suggestion/comment/opinion/reviews about product
or services. There are three types of data commonly found in each
e-commerce site: Review Content, Reviewer and Product Information.

Discussion on review, reviewer and product related attributes regarding
our selected dataset can be seen in Section \ref{subsec:Entities-and-Attributes}.
\textcolor{black}{From the beginning of fake review detection area,
building a dataset for experimentation is a challenging task. Many
researchers forged reviews using different types of sources to build
review dataset }\cite{Algur2010,li2011learning,Ott2011,Wu2010}\textcolor{black}{.
Using forged review dataset (pseudo fake review) is not recommended
by the researchers \cite{Mukherjee2013}. The reason is low classification
accuracy on real life dataset as compared to forged review dataset.
Reviews from real life dataset is considered this research. Real life
dataset is crawled from Yelp \cite{Mukherjee2013a,Zhang2016,Mukherjee2013}.}

\begin{table}[b]
\centering{}\caption{\label{tab:Description-of-Data}Description of Dataset}
\includegraphics[scale=0.4]{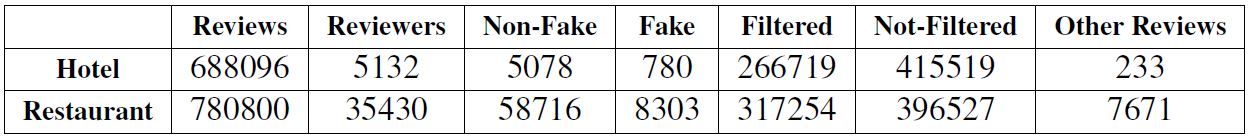}
\end{table}

Yelp was founded in 2004 and the official launch of website in 2005.
The website is source for running businesses where users purchase
products or services and post opinions/reviews/comments on product
and services. In the current position Yelp have more than 1 million
business information, 135+ million reviews and 150+ million distinct
visitor on mobile and web platform. Users can read and post reviews
on businesses. In order to write a review and place a rating (1-5
star), a user must sign in by creating a free account on Yelp that
requires a valid email address. Yelp contains businesses information
of hotels, restaurants, doctor, shopping, automotive (vehicle booking),
beauty parlors, home services, sports and others. \textcolor{black}{Yelp
is popular online review site through which many businesses are getting
clients. Yelp is filtering fake reviews since last decade. Yelp is
using filtering algorithm to detect fake reviews. These techniques
are not known to researchers. Yelp provide filtered dataset that is
available to academic researchers.}

Same review dataset is used by \cite{Mukherjee2013a,Zhang2016} for
experimentation. The dataset used is provided by Dr. Bing Lui. Table
\ref{tab:Description-of-Data} shows number of reviews, reviewers
with respect to hotels and restaurants. Fake reviews are labeled with
``Y'' and non-fake reviews are labeled with ``N''. Whereas ``NR''
and ``YR'' denote filtered and unfiltered reviews.

\subsection{\label{subsec:Entities-and-Attributes}Entities and Attributes}

Two SQlite database is provided that contains crawled review data.
First database contains review data of restaurant and second contains
of hotel. Both database contains three entities in which two are common:
review and reviewer. Resturant database contains three entities: \textit{review},
\textit{reviewer} and \textit{restaurant}. Whereas, hotel database
contains: \textit{review}, \textit{reviewer} and \textit{hotel}. Each
entity and related attributes are defined below:

\subsubsection{{\large{}Restaurant Entity}}

\begin{figure}[H]
\begin{centering}
\includegraphics{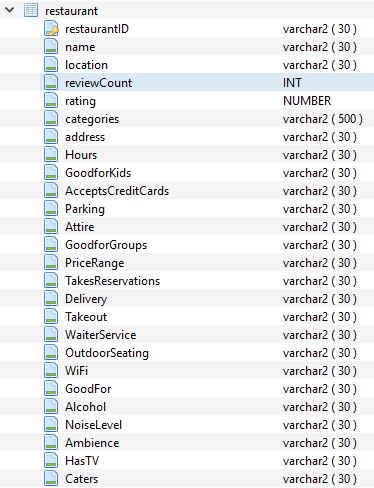}
\par\end{centering}
\centering{}\caption{\label{fig:Restaurant-Entity}Restaurant Entity}
\end{figure}

This entity is contained by restaurant database which consist of metadata
about the restaurant. It contains thirty attributes which contain
details about restaurant. A snapshot of restaurant entity along with
related attributes is shown in Figure \ref{fig:Restaurant-Entity}.
We define some attributes of restaurant entity. However, only \textit{``reviewCount''}
is used in the experimentation. Unique key for identification is stored
in \textit{``restaurantID}''. The attribute ``\textit{name}''
contains name of restaurant and ``\textit{location}'' contains city
and state of restaurant. Attribute ``\textit{reviewCount}'' contains
number of posted reviews on restaurant. Assigned rating of restaurant
by Yelp is stored in ``\textit{rating}''. The attribute ``categories''
contains information about available food items (i.e. Grill Fish,
B.B.Q). In the ``\textit{address}'', full address of restaurant
is given. Opening hours of restaurant are defined in ``\textit{Hours}''.
Suitability of restaurant environment for kids is defined in ``\textit{GoodforKids}''.
The attribute ``\textit{AcceptsCreditCards}'' contains 'Yes' if
credit card is accepted for paying bill otherwise 'No'. Parking facility
in restaurant is notified by ``Parking'' attribute. The attribute
``\textit{Attire}'' contains recommended outfit for customer by
restaurant. Attribute ``\textbf{\textit{GoodforGroups}}'' contains
information that if the environment of restaurant is good for group
meal (e.g. family dinner). The attribute ``PriceRange'' contains
food item rate range. Facility of reservation and food delivery are
defined in ``\textit{TakesReservation}'' and ``Delivery'', respectively. 

\subsubsection{{\large{}Hotel Entity}}

\begin{figure}
\begin{centering}
\includegraphics{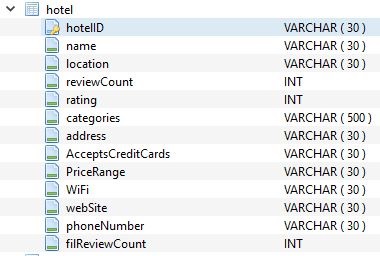}
\par\end{centering}
\caption{Hotel Entity \label{fig:Hotel-Entity}}

\end{figure}

This entity is contained by hotel database which contains thirteen
attributes having details about hotel. A snapshot from database of
hotel entity along with related attributes is shown in Figure \ref{fig:Hotel-Entity}.
We define some attributes of hotel entity. However, only \textit{``reviewCount''}
is used in the experimentation. Unique key for identification is stored
in \textit{``hotelID}''. The attribute ``\textit{name}'' contains
name of the hotel and ``\textit{location}'' contains city and state
of hotel. Attribute ``\textit{reviewCount}'' contains number of
posted reviews on a hotel. Assigned rating of hotel by Yelp is stored
in ``\textit{rating}''. The attribute ``categories'' contains
information about service availability (e.g. event planning, nightlife,
bars, etc). The attribute ``\textit{address}'' contains full address
including street, area, city and state of hotel. 

\subsubsection{{\large{}Reviewer Entity}}

\begin{figure}
\centering{}\includegraphics{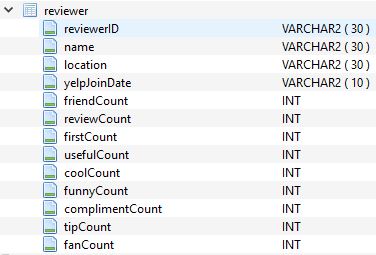}\caption{Reviewer Entity \label{fig:Reviewer-Entity}}
\end{figure}

This entity contains metadata about reviewer and present in restaurant
as well as hotel database. It contains thirteen attributes which contain
details about reviewer profile. Reviewer entity including the following
attributes as shown in Figure \ref{fig:Reviewer-Entity}. We define
some attributes in reviewer entity used in this research. Unique key
for identification is stored in \textit{``reviewerID}''. Account
name of reviewer is stored in ``\textit{name}'' attribute. The attribute
``\textit{location}'' contain residing city of reviewer. Date of
creating account is stored in ``\textit{yelpJoinDate}''. Number
of friends as reviewers are defined in ``\textit{friendCount}''.
The attribute ``\textit{reviewCount}'' contains number of posted
reviews by a reviewer. Total number of useful, cool and funny vote
count on posted reviews of reviewer is defined in ``\textit{usefulCount}'',
``\textit{coolCount}'' and ``\textit{funnyCount}'' respectively. 

\subsubsection{{\large{}Review Entity}}

\begin{figure}[h]
\centering{}\includegraphics{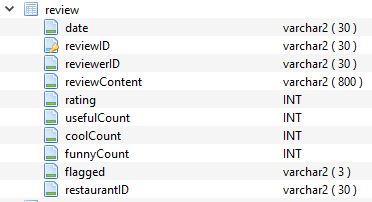}\caption{Review Entity of restaurant \label{fig:Review-Entity}}
\end{figure}

This entity contains metadata about reviewer and present in both restaurant
and hotel database. It contains ten attributes which contain details
about posted reviews. Attributes of review entity in restaurant database
can be seen in Figure \ref{fig:Review-Entity}. 

We define each attribute of review entity. The only difference between
review entity of restaurant and hotel database is that instead of
\textit{``restaurantID}'' hotel database contains \textit{``restaurantID}''.
Unique key for identification of review is stored in \textit{``reviewID}''.
The attribute ``\textit{date}'' contains posted review date. Reviewer
entity and restaurant entity are linked with this entity using ``\textit{reviewerID}''
and ``\textit{restaurantID}''. The textual content of review is
stored in ``\textit{reviewContent}''. Number of useful, cool and
funny vote count over a review are defined in ``\textit{usefulCount}'',
``\textit{coolCount}'' and ``\textit{funnyCount}'' respectively.
Points (rating) on restaurant/hotel given by reviewer is stored in
``rating''. The attribute ``\textit{flagged}'' contains ``Y''
if review is fake otherwise ``N''. 

\section{\label{sec:Preprocessing}Preprocessing}

In many databases of real world contain conflicting and noise data.
The reason is that data is often collected from numerous and heterogeneous
sources. Inconsistency in data results inaccurate outcomes in data
mining process. One of the vital step is preprocessing of data before
initiating process of data mining. There are various preprocessing
methods \cite{sun2007cost} to handle variety of data ( cleansing,
attribute reduction, tokenization, stopwords removing, lemmatization,
and stemming). Two types of preprocessing techniques are used for
this research work: text and data preprocessing. 

\subsection{Text Preprocessing}

Text preprocessing include data mining techniques used to transform
unstructured text. Few text preprocessing techniques on our selected
dataset are defined as follows:
\begin{itemize}
\item \textbf{Tokenization:} Tokenization is task of splitting-up the review
text into words (tokens). i.e. Review content is tokenized into tokens.
For calculating RCS and capital diversity, tokenization is vital step
to separate each word in review.
\item \textbf{Lemmatization:} The task of lemmatizer is to transform word
with respect to morphological root word e.g. 'bought' lemmatized into
'buy'.
\end{itemize}

\subsection{Data Preprocessing}

It is data mining technique to transform raw data into an understandable
data. Data is often noisy, incomplete and/or inconsistent, and may
contain errors. Data preprocessing is a method to resolve such type
of issues. 

It removes unnecessary attributes from review database. In the attribute
\textit{``date''} in \textit{review} entity, the word ``Update''
is concatenated with date (e.g. ``Update - 02-10-2015'' ) to identify
the review that are updated . The term ``Update'' is removed and
attribute data type string is converted into data type date. 

\section{\label{sec:Features}Features Used For Fake Review Detection}

In the projection of this research, two types of features are used:
contextual and behavioral features. For training classification model
contextual and behavioral features are discussed. Our selected predictive
feature set is extracted from Yelp review dataset. However, attributes
used for behavioral features extraction in Yelp dataset may or may
not be available in review data of other e-market sites. 

Contextual and behavioral features are formalized via notations (explained
in Table \ref{tab:List-of-notations}). Each feature either belong
to a review or reviewer denoted by $f(r)$ and $f(a)$ respectively. 

\begin{table}
\centering{}\caption{\label{tab:List-of-notations}List of Associated Notations}
\includegraphics[scale=0.45]{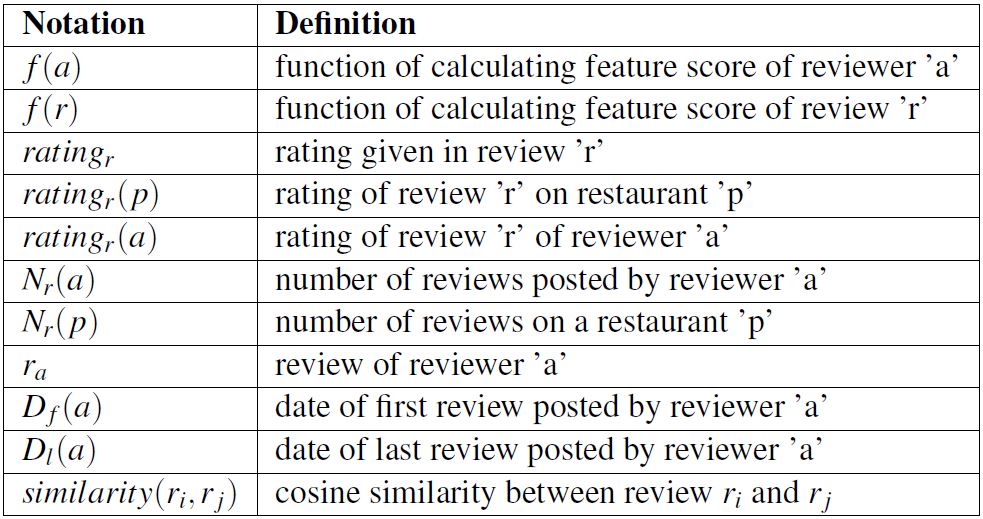}
\end{table}

\subsection{\label{subsec:Contextual-Features}Contextual Features}

Contextual features are also called verbal features are extracted
from review centric feature. Contextual features represent different
perspective of review content of review \cite{Zhang2016}. Many contextual
features are explored which include review length, average review
length, noun ratio, subjectivity, lexical validity, lexical diversity,
capital diversity, sentiment orientation, average content similarity
and others. Average content similarity is adopted in selected feature
set. The reason of selecting average content similarity is that importance
of other contextual features are very less compared with average content
similarity. Average content similarity is also referred as Reviewer
Content Similarity (RCS). RCS is one of the important contextual feature
in this research. It shows average text similarity of all posted reviews
of a reviewer as defined in Equation \ref{eq:RCS}. In RCS, cosine
similarity is used to measure similarity of reviews (defined in Equation
\ref{eq:cosineSimilarity}). Currently researchers are using TFIDF
(term frequency and inverse document frequency) weighting scheme to
weight terms in reviews. TF.IDF is denoted by $LTC$ in smart notations
as defined in Equation \ref{eq:ltc} \cite{salton1988term}. Different
variation of TF.IDF is used to compute RCS. These variation include
$BM25$ and $NNC$(natural term frequency, no document frequency and
cosine) \cite{walker1997some,paltoglou2010study}. Equation \ref{eq:nnc}
and \ref{eq:bm25} define formula of $NNC$ and $BM25$. In defined
equations, $c\left(t,r\right)$ represent numbers of terms ``$t$''
in review content ``$r$''. Here, ``$M$'' denotes number of posted
reviews of reviewer and $df(t)$ denotes number of reviews in which
occurrence of term ``$t$'' is found. 
\begin{equation}
RCS(a)=\frac{\sum_{i}^{n}max(\int_{j}^{n}similarity(r_{i},r_{j}))}{n}\label{eq:RCS}
\end{equation}

\begin{equation}
cosineSimilarity(r_{i},r_{j})=\sum_{k}^{V}r_{i_{k}}.r_{j_{k}}\label{eq:cosineSimilarity}
\end{equation}

\begin{equation}
NNC=\frac{c\left(t_{i},r\right)}{\sqrt{(c(t_{1},r))^{2}+(c(t_{2},r))^{2}\ldots(c(t_{n},r))^{2}}}\label{eq:nnc}
\end{equation}

\begin{equation}
LTC=\frac{\left(1+log\left(c\left(t_{i},r\right)\right)\right).\left(log\left(\frac{N}{df(t)}\right)\right)}{\sqrt{(c(t_{1},r))^{2}+(c(t_{2},r))^{2}\ldots(c(t_{n},r))^{2}}}\label{eq:ltc}
\end{equation}

\begin{equation}
BM25=\frac{c\left(t,r\right).\left(\frac{(k+1).c\left(t,r\right)}{c\left(t,r\right)+k}\right).log\left(\frac{M+1}{df(t)}\right)}{\sqrt{(c(t_{1},r))^{2}+(c(t_{2},r))^{2}\ldots(c(t_{n},r))^{2}}}\label{eq:bm25}
\end{equation}

Another contextual feature named as ``Capital Diversity'' is used
in feature set for hotel dataset. It is the number of capital words
(words starting with capital alphabet) divided by the total number
of token in a review.

\subsection{\label{subsec:Behavioral-Features}Behavioral Features}

Behavioral features are also referred as non-verbal features. These
features capture different behavior of reviewer and its posted reviews.
Many behavioral features are explored by \cite{Zhang2016} for improving
fake review detetion model. We define following behavioral feature
used in this research:
\begin{enumerate}
\item \textbf{Membership Length:} It is defined as number of days between
today and date on which reviewer account was created (see Equation
\ref{eq:member_length}) .
\begin{equation}
M(a)=today-yelpJoinDate(a)\label{eq:member_length}
\end{equation}
\item \textbf{Review Count:} It shows the number of reviews posted by a
reviewer
\item \textbf{Average Posting Rate:} It shows the ratio of total reviews
of a reviewer to number of reviewer active days (see Equation ). An
active day is that on which reviewer has posted atleast one review.
\begin{equation}
APR(a)=\frac{N_{r}(a)}{N(posting\:days)}\label{eq:APR}
\end{equation}
\item \textbf{Positive ratio:} It shows reviews having more than or equal
to 4 as rating value rating divided by total number of reviews of
a reviewer (see Equation \ref{eq:pos-ratio})
\begin{equation}
R_{pos}(a)=\frac{N(\{r_{a}|rating_{r}\geq4\})}{N_{r}(a)}\label{eq:pos-ratio}
\end{equation}
\item \textbf{Positive-to-negative ratio:} It shows the ratio of a reviewer
having more than or equal to 4 reviews rating value to the reviews
having less than or equal to 2 rating value (see Equation \ref{eq:pos-to-neg}).
\begin{equation}
R_{pn}(a)=\frac{N(\{r_{a}|rating_{r}\geq4\})}{N(\{r_{a}|rating_{r}\leq2\})}\label{eq:pos-to-neg}
\end{equation}
\item \textbf{Maximum Posting Rate:} It is the number of maximum posted
reviews in a day (see Equation \ref{eq:MPR}).
\begin{equation}
MPR(a)=Max(\int_{i}^{n}numberofreview())\label{eq:MPR}
\end{equation}
\item \textbf{Review Duration:} Difference of first posted review and last
posted review of reviewer (see Equation \ref{eq:RD-1}). 
\begin{equation}
RD(a)=D_{l}(a)-D_{f}(a)\label{eq:RD-1}
\end{equation}
\item \textbf{Reviewer Deviation: }It captures variation in review rating
on a restaurant. It is computed by subtracting review rating with
absolute deviation of all ratings on a restaurant (see Equation \ref{eq:RevDev}).
\begin{equation}
RevDev(r)=\left|rating-\frac{\sum rating_{r}(p)}{N_{r}(p)}\right|\label{eq:RevDev}
\end{equation}
\end{enumerate}

\section{\label{sec:Classifiers-Adopted}Classifiers Adopted}

On the real life dataset many experiments using various classifiers
is reported including NB, RF, CART, SVM, KNN and others \cite{Zhang2016,Mukherjee2013a,Kaghazgaran2015,Mukherjee2013b}.
We considered SVM and RF classifiers because highest results are reported
using these classifiers on Yelp dataset by \cite{Mukherjee2013a,Zhang2016}.
SMV and RF classifiers are discussed in Section\ref{subsec:Support-Vector-Machine}
and Section\ref{subsec:Random-Forest}. The effect of \textit{``reviewer
deviation''} is explored. The feature \textit{``reviewer deviation''}
is combined with feature set of \cite{Zhang2016}. Review dataset
is scaled for investigating the vitality of new feature set. 

\section{\label{sec:Evaluation-Measures}Evaluation Measures}

In the research area of fake review detection many different evaluation
measures are used to measure accuracy of constructed classification
model. We adopted four predictive accuracy measure for assessing trained
classification model: $Precision$, $Recall$, $F1-measure$ and $Accuracy$.
Formula of these four features are discussed in Section \ref{sec:Classifier-Evaluation}.
We adopted 10-fold cross validation in evaluation process. The average
performance of ten classification results for each classifier is reported.
Most of previous research work on training classifier for fake review
detection adopted 10-fold cross validation. 

\chapter{Experimental Results and Analysis}

\thispagestyle{empty}
\fancyhead[RO]{ EXPERIMENTAL RESULTS AND ANALYSIS }

This chapter discuses a complete experimental setup. This include
selection of reviews and related attributes from real life Yelp database.
Furthermore, extracted contextual and behavioral features from available
attributes of Yelp database is discussed. Main focus was to investigate
behavioral feature ``\textit{reviewer deviation}'' with other behavioral
and contextual features. Further, a contextual feature ``\textit{reviewer
content similarity}'' was explored using different schemes of term
weighting. After preprocessing and computing features, reviews were
classified using RF and SVM classifiers. At last classification models
are evaluated using different evaluation measures. For evaluation
10-fold cross validation is used. We compare classification results
with three perspectives including classifiers, feature sets and term
weighting schemes.

\section{Dataset and Experimental Setup}

First of all, SQlite database of reviews was provided upon making
request to Dr. Bing Liu. Dataset was extracted from these reviews
of Yelp. Further, details of reviews dataset is discussed. Three SQlite
entities are used which are named as ``restaurant/hotel'', ``reviewer''
and ``review''. Each entity consist of several attributes. These
attributes are used in computing contextual and behavioral features.
Twelve contextual and behavioral features are explored including ``reviewer
deviation''. Java is used to extract features from attributes, selecting
dataset based on random reviews, generating CSV files and normalize
data. These extracted features are given to two classifiers RF and
SVM. Both classifiers are implemented in Python. Python provides rich
packages for machine learning. Two popular packages for machine learning
tasks named as `\noun{sklearn}` and '\noun{pandas}'. The result of
classification model is evaluated using precision, recall, F1-measure
and accuracy using 10-fold cross validation. 

Our experimentation consist of a dataset of hotel reviews and two
different sized dataset of restaurant reviews from Yelp database.
Two reviews dataset from restaurant review database were extracted
as shown in Table \ref{tab:Description-of-Data}, we conducted experimentation
on two dataset out of 780800 restaurant reviews. The dataset of hotel
reviews for experimentation can be seen in Table \ref{tab:Dataset-of-Hotel}
in which 1550 random reviews were selected out of 688096 reviews.
The selection of reviews for experiments is based on previously followed
pattern. \textit{Dataset 1} consist of randomly selected 2060 reviews.
In \textit{Dataset 1}, 1964 reviewers posted 2060 reviews. The second
dataset (\textit{Dataset 2}) is scaled upto 12000 reviews. In \textit{Dataset
2}, 9754 reviewers posted on 92 restaurants as shown in Table \ref{tab:Description-of-Data}.
All datasets contain equal number of fake and non-fake reviews. We
selected equal size of fake and non-fake reviews to avoid imbalance
class distribution problem as discussed in section \ref{sec:Imbalance-Class-Distribution}.
The reason behind scaling the review data size is: i) to prove the
vitality of selected feature set, and ii) to report the improvement
in results of evaluation measures. 

\begin{table}[b]
\centering{}\caption{\label{tab:Datasets-restaurant}Datasets of Restaurant}
\begin{tabular}{|c|c|c|c|c|c|}
\hline 
 & \textbf{Restaurants} & \textbf{Reviews} & \textbf{Reviewers} & \textbf{Non-Fake} & \textbf{Fake}\tabularnewline
\hline 
\textbf{Dataset 1} & 31 & 2060 & 1964 & 1030 & 1030\tabularnewline
\hline 
\textbf{Dataset 2} & 92 & 12000 & 9754 & 6000 & 6000\tabularnewline
\hline 
\end{tabular}
\end{table}

\begin{table}
\caption{\label{tab:Dataset-of-Hotel}Dataset of Hotel}

\centering{}%
\begin{tabular}{|c|c|c|c|c|c|}
\hline 
 & \textbf{Hotels} & \textbf{Reviews} & \textbf{Reviewers} & \textbf{Non-Fake} & \textbf{Fake}\tabularnewline
\hline 
\textbf{Dataset 3} & 70 & 1550 & 1499 & 775 & 775\tabularnewline
\hline 
\end{tabular}
\end{table}

Three different combination of contextual and behavioral features
are used in experiments for restaurant datasets. Feature set FS1 and
FS4 consist of eleven features used by \cite{Zhang2016} for restaurant
and hotel reviews respectively. The feature set FS3 consists of features
used by \cite{Mukherjee2013a}. More than thirty features including
both contextual and behavioral features were exploited by \cite{Zhang2016},
but best results were reported using twelve features after feature
pruning. Ten features are common between feature set for hotel reviews
and restaurant reviews. We selected eleven features from feature set
used for restaurant and hotel reviews and added ``Reviewer Deviation''
in each feature set. Our selected dataset for restaurant reviews and
hotel reviews can be seen in Table \ref{tab:Feature-Set-twelve} and
Table \ref{tab:Feature-Set-12-hotel} respectively. The feature set
consists of a behavioral feature ``Reviewer Deviation'' along with
feature used by \cite{Zhang2016} for restaurant reviews and referred
as FS2. A feature set consists of a behavioral feature ``Reviewer
Deviation'' along with eleven feature used by \cite{Zhang2016} for
hotel reviews and referred as FS5. Our feature set consist of twelve
feature including one contextual and eleven behavioral features for
restaurant reviews as shown in Table \ref{tab:Feature-Set-twelve}.
Second feature set for hotel reviews include two contextual and ten
behavioral features as shown in Table \ref{tab:Feature-Set-12-hotel}.
Another feature set used by \cite{Mukherjee2013a} consist of six
features including four behavioral features, a contextual feature
and unigram (shown in Table \ref{tab:Feature-Set-five}). 

\begin{table}
\centering{}\caption{\label{tab:Feature-Set-11}Features Used by \textit{Zhang} \textit{et.al}
for restaurant reviews}
\begin{tabular}{|c|c|}
\hline 
 & \textbf{Features (FS1)}\tabularnewline
\hline 
1. & Useful Count\tabularnewline
\hline 
2. & Cool Count\tabularnewline
\hline 
3. & Funny Count\tabularnewline
\hline 
4. & Friend Count\tabularnewline
\hline 
5. & Review Count\tabularnewline
\hline 
6. & Average Posting Rate\tabularnewline
\hline 
7. & Positive Ratio\tabularnewline
\hline 
8. & Reviewer Content Similarity\tabularnewline
\hline 
9. & Membership Length\tabularnewline
\hline 
10 & Review Duration\tabularnewline
\hline 
11. & Positive to Negative Ratio\tabularnewline
\hline 
\end{tabular}
\end{table}

\begin{table}
\centering{}\caption{Feature Set of Reviewer Deviation and Other Features used by \textit{Zhang}
\textit{et.al}\textbf{ }for restaurant reviews\label{tab:Feature-Set-twelve}}
\begin{tabular}{|c|c|}
\hline 
 & \textbf{Features (FS2)}\tabularnewline
\hline 
1. & Useful Count\tabularnewline
\hline 
2. & Cool Count\tabularnewline
\hline 
3. & Funny Count\tabularnewline
\hline 
4. & Friend Count\tabularnewline
\hline 
5. & Review Count\tabularnewline
\hline 
6. & Average Posting Rate\tabularnewline
\hline 
7. & Positive Ratio\tabularnewline
\hline 
8. & Reviewer Content Similarity\tabularnewline
\hline 
9. & Membership Length\tabularnewline
\hline 
10 & Review Duration\tabularnewline
\hline 
11. & Positive to Negative Ratio\tabularnewline
\hline 
12. & Reviewer Deviation\tabularnewline
\hline 
\end{tabular}
\end{table}

\begin{table}
\centering{}\caption{Feature Set Used by \textit{Mukherjee et.al} for Yelp reviews\label{tab:Feature-Set-five} }
\begin{tabular}{|c|c|}
\hline 
 & \textbf{Features (FS3)}\tabularnewline
\hline 
1. & Content Length\tabularnewline
\hline 
2. & Positive Ratio\tabularnewline
\hline 
3. & Reviewer Content Similarity\tabularnewline
\hline 
4. & Reviewer Deviation\tabularnewline
\hline 
5. & Maximum Number of Reviews\tabularnewline
\hline 
6. & Unigrams\tabularnewline
\hline 
\end{tabular}
\end{table}

\begin{table}
\centering{}\caption{\label{tab:Feature-Set-11-hotel}Feature Used by \textit{Zhang} \textit{et.al}
for hotel reviews}
\begin{tabular}{|c|c|}
\hline 
 & \textbf{Features (FS4)}\tabularnewline
\hline 
1. & Useful Count\tabularnewline
\hline 
2. & Cool Count\tabularnewline
\hline 
3. & Funny Count\tabularnewline
\hline 
4. & Friend Count\tabularnewline
\hline 
5. & Review Count\tabularnewline
\hline 
6. & Average Posting Rate\tabularnewline
\hline 
7. & Tips Count\tabularnewline
\hline 
8. & Reviewer Content Similarity\tabularnewline
\hline 
9. & Membership Length\tabularnewline
\hline 
10 & Review Duration\tabularnewline
\hline 
11. & Capital Diversity\tabularnewline
\hline 
\end{tabular}
\end{table}

\begin{table}
\centering{}\caption{\label{tab:Feature-Set-12-hotel}Feature Set of Reviewer Deviation
and Other Features used by \textit{Zhang} \textit{et.al}\textbf{ }for
hotel reviews}
\begin{tabular}{|c|c|}
\hline 
 & \textbf{Features (FS5)}\tabularnewline
\hline 
1. & Useful Count\tabularnewline
\hline 
2. & Cool Count\tabularnewline
\hline 
3. & Funny Count\tabularnewline
\hline 
4. & Friend Count\tabularnewline
\hline 
5. & Review Count\tabularnewline
\hline 
6. & Average Posting Rate\tabularnewline
\hline 
7. & Tips Count\tabularnewline
\hline 
8. & Reviewer Content Similarity\tabularnewline
\hline 
9. & Membership Length\tabularnewline
\hline 
10 & Review Duration\tabularnewline
\hline 
11. & Capital Diversity\tabularnewline
\hline 
12. & Reviewer Deviation\tabularnewline
\hline 
\end{tabular}
\end{table}

\begin{figure}[h]
\begin{centering}
\includegraphics[width=5in]{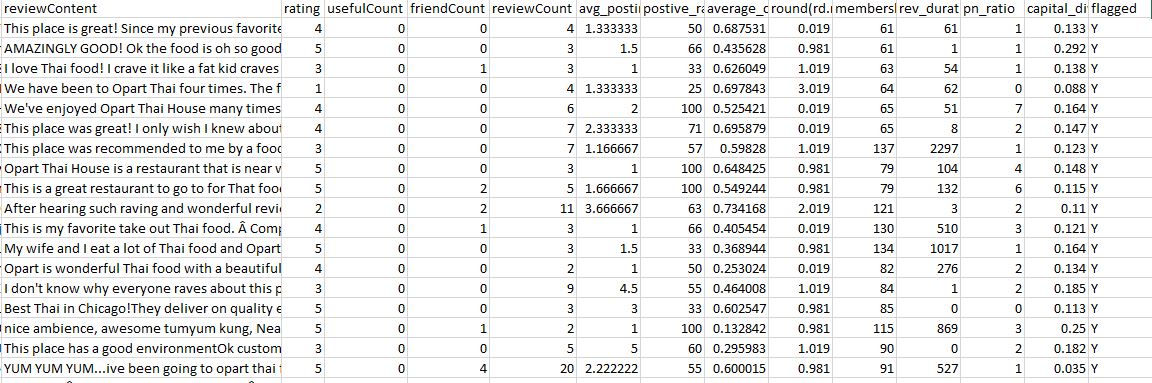}\caption{Snapshot of dataset \label{fig:Snapshot-of-dataset}}
\par\end{centering}
\end{figure}

A snapshot of a dataset is shown in Figure \ref{fig:Snapshot-of-dataset}
where each row represents a review. Each column in Figure \ref{fig:Snapshot-of-dataset}
represent a feature. RCS is one of the important contextual feature.
RCS shows average text similarity of all posted reviews of a reviewer.
RCS is discussed in Section \ref{subsec:Behavioral-Features}. In
RCS, cosine similarity is used to measure similarity of reviews. Currently
researchers are using cosine similarity based on TF.IDF weighting
scheme to weight terms in reviews. Different variation of TF.IDF is
used to compute RCS. These variation include BM25 and NNC. 

In short, one of the focus was to extract data from real life dataset.
Three datasets as mentioned earlier were extracted from Yelp database
to investigate the effect of behavioral feature ``\textit{reviewer
deviation}'' combined with other contextual and behavioral features
on classification of fake reviews. Another focus was to explore a
contextual feature RCS using different variations of weighting terms
in reviews for computation of text similarity. SVM and RF is used
for classification . Ten fold cross validation is used for evaluation.
Different evaluation measures accuracy, precision, recall and F1-measure
are used. 

\section{\label{sec:Experimental-Results}Importance of Reviewer Deviation}

Experiments were made on three datasets, two dataset consist of 2060
and 12000 reviews on restaurants and third dataset consist of 1550
reviews on hotels. Datasets were extracted from Yelp dataset which
considered to be real life dataset. Three different feature sets were
used for restaurant reviews and two feature sets for hotel reviews.
Three of these feature sets for restaurant and hotel reviews are used
by \cite{Zhang2016} and \cite{Mukherjee2013}. One of the feature
set comprises of a behavioral feature named as ``\textit{Reviewer
Deviation}'' and eleven features used by \cite{Zhang2016} for restaurant
and hotel reviews. One of the focus of this research was to investigate
the importance of ``\textit{Reviewer Deviation}'' along with contextual
and behavioral feature used by \cite{Zhang2016}. Interestingly, it
was found after computing importance score that the importance score
of ``\textit{Reviewer Deviation}'' is ranked in top ten contextual
and behavioral features. Results on using ``\textit{Reviewer Deviation}''
along with other contextual and behavioral features used by \cite{Zhang2016}
and \cite{Mukherjee2013}, shows improvement in terms of precision
and accuracy. Detailed analysis is made on following section.

\subsection{Importance Score of Contextual and Behavioral Features}

The importance score of each contextual and behavioral features on
dataset 1 with FS2 and dataset 3 with FS5 are computed. Behavioral
feature ``reviewer deviation'' is ranked among top ten features
according to computation of importance score for ``reviewer deviation''
using RF as shown in Table \ref{tab:Importance-Score-of} and Table
\ref{tab:Importance-Score-of-Hotel}. The reason for using RF for
computation of importance score of each feature is that it outperform
SVM. 

Dataset 1 contains reviews from Yelp restaurant database. Feature
ranking on Dataset 1 shows that useful, cool, funny votes are most
important features for restaurant reviews. Whereas number of posted
reviews and friend of reviewer acquired low ranking as to the above
mentioned features. Usefulness of votes count show that other users
also support review content. Another behavior of spammer is identified
by fifth feature which defines that total number of posted reviews
by spammer is much greater than a true reviewer. ``\textit{Average
posting rate}'' ranked sixth proves that considering posting frequency
of a reviewer is also a vital perspective for fake review detection.
It is analyzed by many researchers that 80\% of reviews by spammers
deviates towards positive polarity \cite{li2011learning,algur2010spam}.
Our observations also support previous reported results about positive
ratio which is on seventh rank. Initial study on fake review detection
by \cite{Jindal2007,Jindal2007a} were based on content similarity
of reviews and similarity of 90\% reviews with content of other reviews
were considered as fake. Likewise top eighth feature which relates
to contextual feature is proved important which support observation
that reviews of spammer have textual resemblance with each other \cite{Jindal2007,Zhang2016,Mukherjee2013,MyleOtt,Ott2011,LinY2014}.
Deviation of rating of a review from other reviews on a restaurant
increase the change of ``spamicity''. That is why \textit{``reviewer
deviation}'' is observed on ninth rank which support the research
investigation objective for empirically proving vitality of this behavioral
feature. ``\textit{Membership Length}'' is investigated due to assumption
in literature that more old the reviewer's account is more reliable
it is. In some cases spammer create account to place false reviews
for a little time period. This behavior of spammer is captured with
'\textit{Review Duration}' which is ranked eleventh. The last behavior
is the ratio between positive and negative rating of reviews by reviewer.
It is interpreted that variation in the ratio of positive reviews
compared with negative reviews of fake reviewer is different than
true reviewer \cite{Zhang2016}. 

Dataset 2 contains reviews from Yelp hotel database and feature ranking
for classification can be seen in Table \ref{tab:Importance-Score-of-Hotel}.
Feature ranking on Dataset 2 shows that ``\textit{Review Duration}'',
\textit{``Tips Count}'' and ``\textit{Useful} \textit{Count}''
obtained highest score among all features. Whereas, \textit{``Capital
Diversity}'', \textit{``Cool Count}'' and \textit{``Funny Count}''
obtained lowest importance score in classification. The contextual
feature \textit{``Reviewer Cotent Similarity}'' is more effective
for classifying hotel reviews hence it gained fifth ranking. The feature
under consideration \textit{``reviewer deviation}'' is observed
on eighth rank which support the research investigation objective
for empirically proving vitality of this behavioral feature. 

The focus was to investigate the importance of ``\textit{Reviewer
deviation}'' in combination with contextual and behavioral features
used by \cite{Zhang2016} and \cite{Mukherjee2013}. This feature
was compared with feature sets used by \cite{Zhang2016} and \cite{Mukherjee2013}.
\cite{Zhang2016} and \cite{Mukherjee2013} used a contextual feature
``\textit{Reviewer Content Similarity}'' based on TF.IDF (LTC).
Experiments were made using RF and SVM. For evaluation 10-fold cross
validation is used. Three datasets were used that include restaurant
and hotel reviews. Dataset 1 and Dataset 2 consist of 2060 and 12000
restaurant reviews respectively as shown in Table \ref{tab:Datasets-restaurant}.
Dataset 3 consists of 1550 hotel reviews as shown in Table \ref{tab:Dataset-of-Hotel}.

\begin{table}[h]
\centering{}\caption{Importance Score of Selected Set of Features on Restaurant Reviews
\label{tab:Importance-Score-of}}
\begin{tabular}{clc}
\hline 
\textbf{Rank} & \textbf{Features} & \textbf{Importance Score}\tabularnewline
\hline 
\hline 
1 & Useful Count & 24.668\tabularnewline
2 & Cool Count & 18.649\tabularnewline
3 & Funny Count & 14.638\tabularnewline
4 & Friend Count & 12.239\tabularnewline
5 & Review Count & 9.017\tabularnewline
6 & Average Posting Rate & 5.031\tabularnewline
7 & Positive Ratio & 4.222\tabularnewline
8 & Reviewer Content Similarity & 3.408\tabularnewline
9 & Reviewer Deviation & 2.993\tabularnewline
10 & Membership Length & 2.504\tabularnewline
11 & Review Duration & 1.787\tabularnewline
12 & Positive Negative Ratio & 0.843\tabularnewline
\hline 
\end{tabular}
\end{table}

\begin{table}[h]
\caption{Importance Score of Selected Set of Features on Hotel Reviews \label{tab:Importance-Score-of-Hotel}}

\centering{}%
\begin{tabular}{clc}
\hline 
\textbf{Rank} & \textbf{Features} & \textbf{Importance Score}\tabularnewline
\hline 
\hline 
1 & Review Duration & 23.934\tabularnewline
2 & Tips Count & 21.082\tabularnewline
3 & Useful Count & 13.545\tabularnewline
4 & Review Count & 12.323\tabularnewline
5 & Reviewer Content Similarity & 10.871\tabularnewline
6 & Average Posting Rate & 4.260\tabularnewline
7 & Membership Length & 3.432\tabularnewline
8 & Reviewer Deviation & 2.940\tabularnewline
9 & Friend Count & 2.593\tabularnewline
10 & Capital Diversity & 2.353\tabularnewline
11 & Cool Count & 1.668\tabularnewline
12 & Funny Count & 0.993\tabularnewline
\hline 
\end{tabular}
\end{table}

\subsection{Results and Analysis on Dataset 1}

\begin{figure}[h]
\begin{centering}
\includegraphics[scale=0.5]{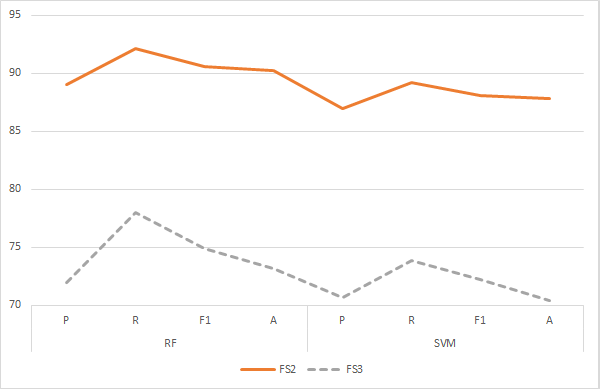}
\par\end{centering}
\caption{Comparison of FS2 and FS3 on Dataset 1 \label{fig:Comparison-of-FS2-Vs-FS3}}
\end{figure}

FS3 achieved precision, recall, f1-score and accuracy of 71.96\%,
77.96\%, 74.84\% and 73.17\% respectively on Dataset 1 using RF classifier
as shown in Table \ref{tab:Results-dataset1-ltc}. Whereas FS2 achieved
precision, recall, f1-score and accuracy of 89.07\%, 92.13\%, 90.57\%
and 90.23\% respectively. The improvement here in accuracy, precision,
recall and f2-score is 17.06\%, 17.1\%, 14.17\% and 15.73\% respectively.
This improvement is quite significant. Similarity, using SVM the accuracy,
precision, recall and f1-score on FS3 is 70.43\%, 70.64\%, 73.88\%
and 72.22\% respectively. Whereas FS2 gives accuracy, precision, recall
and f1-measure of 87.86\%, 87\%, 89.22\% and 88.09\% respectively.
It gives improvement of 17.42\%, 16.36\%, 15.34\% and 15.87\% in terms
of accuracy, precision, recall and f1-measure respectively which is
quite significant. The visualization of FS2 and FS3 can be seen in
Figure \ref{fig:Comparison-of-FS2-Vs-FS3}.

\begin{figure}[h]
\begin{centering}
\includegraphics[scale=0.5]{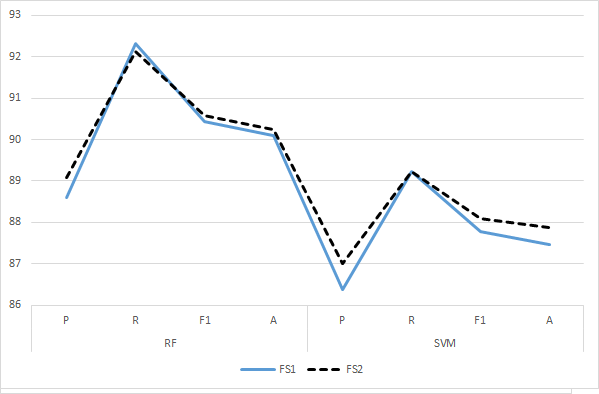}
\par\end{centering}
\caption{Comparison of FS1 and FS2 on Dataset 1 \label{fig:Comparison-of-FS1-Vs-FS2}}

\end{figure}

FS1 achieved accuracy, precision, recall and f1-measure of 90.09\%,
88.6\%, 92.33\%, 90.42\% respectively using RF on dataset1 as shown
in Table \ref{tab:Results-dataset1-ltc}. The improvement of FS2 compared
with FS1 is 0.47\%, 0.15\%, 0.13\% in terms of precision, recall,
f1-measure and accuracy respectively using RF. Whereas improvement
using SVM is 0.63\%, 0.32\%, 0.38\% in terms of precision, f1-measure
and accuracy respectively. Improvements using both classifiers are
quite significant. The visualization of FS1 and FS2 can be seen in
Figure \ref{fig:Comparison-of-FS1-Vs-FS2}.

\begin{table}[h]
\centering{}\caption{\label{tab:Results-dataset1-ltc}Results on Dataset 1 Using LTC}
\begin{tabular}{|c|c|c|c|c|c|}
\hline 
\textbf{Classifier} & \textbf{Feature Set} & \textbf{Precision} & \textbf{Recall} & \textbf{F1} & \textbf{Accuracy}\tabularnewline
\hline 
 & FS1 & 88.600 & 92.330 & 90.426 & 90.097\tabularnewline
\cline{2-6} \cline{3-6} \cline{4-6} \cline{5-6} \cline{6-6} 
RF & FS2 & 89.073 & 92.135 & 90.578 & \textbf{90.231}\tabularnewline
\cline{2-6} \cline{3-6} \cline{4-6} \cline{5-6} \cline{6-6} 
 & FS3 & 71.967 & 77.961 & 74.844 & 73.170\tabularnewline
\hline 
 & FS1 & 86.370 & 89.223 & 87.773 & 87.475\tabularnewline
\cline{2-6} \cline{3-6} \cline{4-6} \cline{5-6} \cline{6-6} 
SVM & FS2 & 87.002 & 89.223 & 88.098 & 87.864\tabularnewline
\cline{2-6} \cline{3-6} \cline{4-6} \cline{5-6} \cline{6-6} 
 & FS3 & 70.641 & 73.883 & 72.226 & 70.439\tabularnewline
\hline 
\end{tabular}
\end{table}

\subsection{Results and Analysis on Dataset 2}

\begin{figure}[h]
\begin{centering}
\includegraphics[scale=0.5]{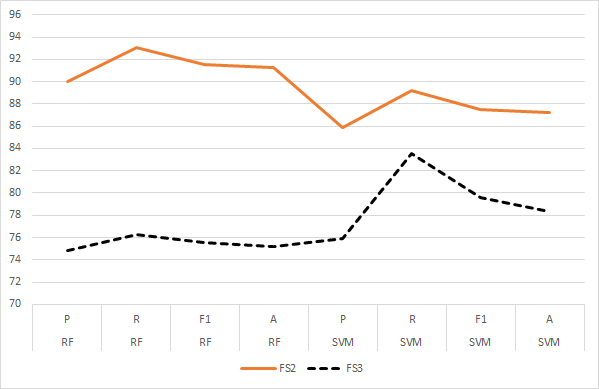}\caption{Comparison of FS2 and FS3 on Dataset 2 \label{fig:Comparison-of-FS2-FS3-Dataset2}}
\par\end{centering}
\end{figure}

FS3 achieved precision, recall, f1-score and accuracy of 74.86\%,
76.29\%, 75.57\% and 75.18\% respectively on Dataset 2 using RF classifier
as shown in Table \ref{tab:Experiment-dataset2-LTC}. Whereas FS2
achieved precision, recall, f1-score and accuracy of 90.01\%, 93.03\%,
91.49\% and 91.24\% respectively. The significant improvement here
in precision, recall, f1-measure and accuracy is 15.15\%, 16.73\%,
15.92\% and 16.06\% respectively. Similarity, using SVM the precision,
recall, f1-measure and accuracy on FS3 is 75.95\%, 83.59\%, 79.58\%
and 78.37\% respectively. Whereas FS2 gives precision, recall, f1-measure
and accuracy of 85.90\%, 89.21\%, 87.52\% and 87.22\% respectively.
It gives improvement of 9.95\%, 5.62\%, 7.94\%, 8.85\% in terms of
precision, recall, f1-measure and accuracy respectively which is quite
significant. Comparison between FS2 and FS3 is visualized in Figure
\ref{fig:Comparison-of-FS2-FS3-Dataset2}. 

\begin{table}[h]
\centering{}\caption{Results on Dataset 2 Using LTC \label{tab:Experiment-dataset2-LTC}}
\begin{tabular}{|c|c|c|c|c|c|}
\hline 
\textbf{Classifier} & \textbf{Feature Set} & \textbf{Precision} & \textbf{Recall} & \textbf{F1} & \textbf{Accuracy}\tabularnewline
\hline 
 & FS1 & 90.545 & 92.983 & 91.748 & 91.172\tabularnewline
\cline{2-6} \cline{3-6} \cline{4-6} \cline{5-6} \cline{6-6} 
RF & FS2 & 90.015 & 93.032 & 91.499 & \textbf{91.244}\tabularnewline
\cline{2-6} \cline{3-6} \cline{4-6} \cline{5-6} \cline{6-6} 
 & FS3 & 74.862 & 76.295 & 75.572 & 75.184\tabularnewline
\hline 
 & FS1 & 85.643 & 88.737 & 87.163 & 86.868\tabularnewline
\cline{2-6} \cline{3-6} \cline{4-6} \cline{5-6} \cline{6-6} 
SVM & FS2 & 85.907 & 89.213 & 87.529 & 87.229\tabularnewline
\cline{2-6} \cline{3-6} \cline{4-6} \cline{5-6} \cline{6-6} 
 & FS3 & 75.951 & 83.590 & 79.588 & 78.375\tabularnewline
\hline 
\end{tabular}
\end{table}

\begin{figure}[h]
\centering{}\includegraphics[scale=0.5]{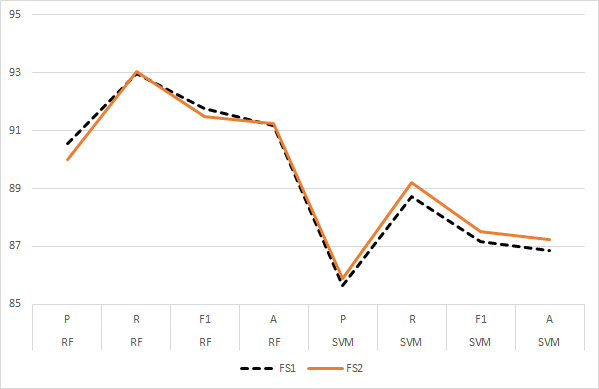}\caption{Comparison of FS1 and FS2 on Dataset 2 \label{fig:Comparison-of-FS1-FS2-Dataset2}}
\end{figure}

FS1 achieved precision, recall, f1-measure and accuracy of 90.54\%,
92.98\%, 91.74\% and 91.17\% respectively using RF on dataset1 as
shown in Table \ref{tab:Results-dataset1-ltc}. The improvement of
FS2 compared with FS1 is 0.05\% and 0.07\% in terms of recall and
accuracy respectively using RF. Whereas improvement using SVM is 0.26\%,
0.48\%, 0.37\% and 0.36\% in terms of precision, recall, f1-measure
and accuracy respectively. In comparison with Dataset 1 improvement
of 1\% in accuracy and F1 is observed. The results of FS1 and FS2
are visualized in Figure \ref{fig:Comparison-of-FS1-FS2-Dataset2}.

\subsection{Results and Analysis on Dataset 3}

\begin{figure}[h]
\centering{}\includegraphics[scale=0.5]{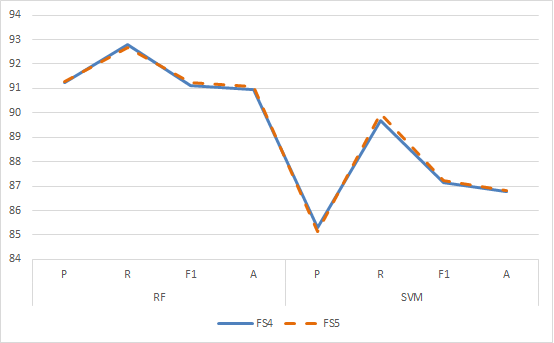}\caption{Comparison of FS4 and FS5 on Dataset 3 \label{fig:Comparison-of-FS4-FS5-Dataset3}}
\end{figure}

The results of RF on Dataset 3 using FS5 acquired highest accuracy
of 91.096\% which is observed due to increase in recall as shown in
Table \ref{tab:Results-on-Dataset3}. Whereas, FS4 obtained 90.967\%
accuracy which is the little difference between the accuracy with
FS5. Similarity, using SVM the precision, recall, f1-measure and accuracy
on FS5 is 85.15\%, 89.84\%, 87.28\% and 86.83\% respectively. This
difference between both classification on FS4 and FS5 shows small
increase in recall. The comparison between results of both feature
set with classification results can be visualized in Figure \ref{fig:Comparison-of-FS4-FS5-Dataset3}.

\begin{table}
\caption{Results on Dataset 3 Using LTC \label{tab:Results-on-Dataset3}}

\begin{centering}
\begin{tabular}{|c|c|c|c|c|c|}
\hline 
\textbf{Classifier} & \textbf{Feature Set} & \textbf{Precision} & \textbf{Recall} & \textbf{F1} & \textbf{Accuracy}\tabularnewline
\hline 
\multirow{2}{*}{RF} & FS4 & 91.226 & 92.782 & 91.102 & 90.967\tabularnewline
\cline{2-6} \cline{3-6} \cline{4-6} \cline{5-6} \cline{6-6} 
 & FS5 & 91.155 & 92.925 & 91.269 & \textbf{91.096}\tabularnewline
\hline 
\multirow{2}{*}{SVM} & FS4 & 85.301 & 89.696 & 87.129 & 86.774\tabularnewline
\cline{2-6} \cline{3-6} \cline{4-6} \cline{5-6} \cline{6-6} 
 & FS5 & 85.153 & 89.840 & 87.287 & 86.838\tabularnewline
\hline 
\end{tabular}
\par\end{centering}
\end{table}

\section{RCS Based on Different Weighting Schemes }

The most important contextual feature ``\textit{Reviewer Content
Similarity}'' (RCS) which calculates similarity between reviews of
a reviewer to identify duplicates or partial duplicates reviews. This
contextual feature RCS is used along with other behavioral features.
Focus was to investigate RCS using different term weighting schemes.
In literature $LTC$ (TF.IDF) term weighting scheme is used in calculating
cosine similarity between two reviews. Two term weighting schemes
are explored: $NNC$ and $BM25$. Formal representation of these weighting
schemes are discussed in Section \ref{subsec:Contextual-Features}.
Main focus was to investigate the effect of these weighting schemes
on RCS. Experiments were carried out to explore if RCS based on these
weighting schemes have any effect on fake review detection. Experiments
were carried out to detect fake reviews based on RCS (using different
term weighting schemes) along with other behavioral features. RF and
SVM were used as classifiers and 10-fold cross validation is used
for evaluation. Experiments were carried out on two datasets. Detailed
analysis on results is given in the following section.

\subsection{Results and Analysis on Dataset 1}

The results of RCS based on NNC along with other behavioral features
can be seen in Table \ref{tab:Results-dataset1-nnc}. FS2 achieved
precision, recall, f1-measure and accuracy of 89.55\%, 92.13\%, 90.82\%
and 90.19\% respectively using RF on Dataset 1. The improvement of
FS2 compared with FS3 is 17.59\%, 14.17\%, 15.98\% and 17.02\% in
terms of precision, recall, f1-measure and accuracy respectively.
Whereas comparing with FS1 shows improvement of precision, f1-measure
and accuracy of 0.96\%, 0.40\% and 0.1\% respectively. Similarly using
SVM, FS2 achieved precision, recall, f1-measure and accuracy of 83.72\%,
87.37\%, 85.51\% and 85.09\% respectively which shows improvement
of 13.08\%, 13.49\%, 13.28\% and 14.65\% in terms of precision, recall,
f1-measure and accuracy compared with FS1. 

\begin{table*}
\centering{}\caption{Results of Dataset 1 Using NNC \label{tab:Results-dataset1-nnc}}
\begin{tabular}{|c|c|c|c|c|c|c|}
\hline 
\textbf{Classifier} & \textbf{Feature Set} & \textbf{Precision} & \textbf{Recall} & \textbf{F1} & \textbf{Accuracy} & \textbf{Weighting Scheme}\tabularnewline
\hline 
 & FS1 & 88.600 & 92.330 & 90.426 & 90.097 & LTC\tabularnewline
\cline{2-7} \cline{3-7} \cline{4-7} \cline{5-7} \cline{6-7} \cline{7-7} 
RF & FS2 & 89.559 & 92.135 & 90.829 & 90.194 & NNC\tabularnewline
\cline{2-7} \cline{3-7} \cline{4-7} \cline{5-7} \cline{6-7} \cline{7-7} 
 & FS3 & 71.967 & 77.961 & 74.844 & 73.170 & LTC\tabularnewline
\hline 
 & FS1 & 86.370 & 89.223 & 87.773 & 87.475 & LTC\tabularnewline
\cline{2-7} \cline{3-7} \cline{4-7} \cline{5-7} \cline{6-7} \cline{7-7} 
SVM & FS2 & 83.727 & 87.378 & 85.513 & 85.097 & NNC\tabularnewline
\cline{2-7} \cline{3-7} \cline{4-7} \cline{5-7} \cline{6-7} \cline{7-7} 
 & FS3 & 70.641 & 73.883 & 72.226 & 70.439 & LTC\tabularnewline
\hline 
\end{tabular}
\end{table*}

The results of RCS based on $BM25$ along with other behavioral features
can be seen in Table \ref{tab:Results-dataset1-BM}. FS2 achieved
precision, recall, f1-measure and accuracy of 89.55\%, 92.13\%, 90.82\%
and 90.19\% respectively using RF. Compared with FS3, the improvement
of 17.59\%, 14.17\%, 15.98\% and 17.02\% in terms of precision, recall,
f1-measure and accuracy respectively. Whereas improvement of 0.96\%,
0.40\%, 0.1\% in precision, f1-measure and accuracy respectively is
observed. Similarity using SVM, FS2 achieved precision, recall, f1-measure
and accuracy of 83.72\%, 87.37\%, 85.51\% and 85.09\% respectively.
The improvement compared with FS3 is 13.086\%, 13.495, 13.28\% and
14.65\% in terms of precision, recall, F1-measure and accuracy respectively. 

\begin{table}[H]
\centering{}\caption{Results of Dataset 1 Using BM25 \label{tab:Results-dataset1-BM}}
\begin{tabular}{|c|c|c|c|c|c|c|}
\hline 
\textbf{Classifier} & \textbf{Feature Set} & \textbf{Precision} & \textbf{Recall} & \textbf{F1} & \textbf{Accuracy} & \textbf{Weighting Scheme}\tabularnewline
\hline 
 & FS1 & 88.600 & 92.330 & 90.426 & 90.097 & LTC\tabularnewline
\cline{2-7} \cline{3-7} \cline{4-7} \cline{5-7} \cline{6-7} \cline{7-7} 
RF & FS2 & 90.033 & 92.233 & 91.119 & \textbf{90.776} & BM25\tabularnewline
\cline{2-7} \cline{3-7} \cline{4-7} \cline{5-7} \cline{6-7} \cline{7-7} 
 & FS3 & 71.967 & 77.961 & 74.844 & 73.170 & LTC\tabularnewline
\hline 
 & FS1 & 86.370 & 89.223 & 87.773 & 87.475 & LTC\tabularnewline
\cline{2-7} \cline{3-7} \cline{4-7} \cline{5-7} \cline{6-7} \cline{7-7} 
SVM & FS2 & 86.978 & 89.611 & 88.275 & 88.009 & BM25\tabularnewline
\cline{2-7} \cline{3-7} \cline{4-7} \cline{5-7} \cline{6-7} \cline{7-7} 
 & FS3 & 70.641 & 73.883 & 72.226 & 70.439 & LTC\tabularnewline
\hline 
\end{tabular}
\end{table}

\begin{figure}
\begin{centering}
\includegraphics[width=6in]{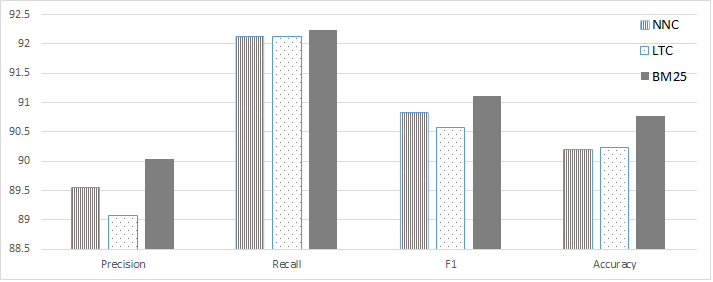}
\par\end{centering}
\caption{Results of NNC, LTC and BM25 Schemes of FS2 on Dataset 1 \label{fig:Result-Comparison-with}}
\end{figure}

The result comparison of all three term weighting schemes of FS2 using
RF are visualized in Figure \ref{fig:Result-Comparison-with} of Dataset
1 from which we can justify that $BM25$ weighting scheme improves
precision, f1-score and accuracy. 

\subsection{Results and Analysis on Dataset 2}

The results of RCS based on NNC along with other behavioral features
can be seen in Table \ref{tab:Results-Dataset2-NNC}. FS2 achieved
precision, recall, f1-measure and accuracy of 89.32\%, 92.81\%, 91.03\%
and 91.08\% respectively using RF on Dataset 1. This results does
not show improvement compared with FS1. However, the improvement of
FS2 compared with FS3 is 14.46\%, 16.52\%, 15.46\% and 15.89\% in
terms of precision, recall, f1-measure and accuracy respectively.
Similarly using SVM, FS2 achieved precision, recall, f1-measure and
accuracy of 84.80\%, 90.04\% , 87.34\% and 86.86\% respectively which
shows improvement of 8.85\%, 6.459\%, 7.7\% and 8.48\% in terms of
precision, recall, f1-measure and accuracy compared with FS3. 

\begin{table*}
\centering{}\caption{Results of Dataset 2 with NNC \label{tab:Results-Dataset2-NNC} }
\begin{tabular}{|c|c|c|c|c|c|c|}
\hline 
\textbf{Classifier} & \textbf{Feature Set} & \textbf{Precision} & \textbf{Recall} & \textbf{F1} & \textbf{Accuracy} & \textbf{Weighting Scheme}\tabularnewline
\hline 
 & FS1 & 90.545 & 92.983 & 91.748 & 91.172 & LTC\tabularnewline
\cline{2-7} \cline{3-7} \cline{4-7} \cline{5-7} \cline{6-7} \cline{7-7} 
RF & FS2 & 89.322 & 92.819 & 91.037 & 91.081 & NNC\tabularnewline
\cline{2-7} \cline{3-7} \cline{4-7} \cline{5-7} \cline{6-7} \cline{7-7} 
 & FS3 & 74.862 & 76.295 & 75.572 & 75.184 & LTC\tabularnewline
\hline 
 & FS1 & 85.643 & 88.737 & 87.163 & 86.868 & LTC\tabularnewline
\cline{2-7} \cline{3-7} \cline{4-7} \cline{5-7} \cline{6-7} \cline{7-7} 
SVM & FS2 & 84.806 & 90.049 & 87.349 & 86.860 & NNC\tabularnewline
\cline{2-7} \cline{3-7} \cline{4-7} \cline{5-7} \cline{6-7} \cline{7-7} 
 & FS3 & 75.951 & 83.590 & 79.588 & 78.375 & LTC\tabularnewline
\hline 
\end{tabular}
\end{table*}

The results of RCS based on $BM25$ along with other behavioral features
can be seen in Table \ref{tab:Results-Dataset2-BM}. FS2 achieved
precision, recall, f1-measure and accuracy of 90.67\%, 92.95\%, 91.86\%
and 91.39\% respectively using RF. Compared with FS3, the improvement
of 15.81\%, 16.65\%, 16.29\% and 16.21\% in terms of precision, recall,
f1-measure and accuracy respectively. Whereas improvement of 0.13\%,0.11\%
and 0.22\% in precision, f1-measure and accuracy respectively is observed.
Similarity using SVM with FS2 achieved precision, recall, f1-measure
and accuracy of 85.90\%, 89.19\%, 87.51\% and 87.22\% respectively.
The improvement compared with FS1 is 0.26\%, 0.46\%, 0.36\% and 0.35\%
in terms of precision, recall, F1-measure and accuracy respectively. 

\begin{table*}
\centering{}\caption{Results of Dataset 2 Using BM25 \label{tab:Results-Dataset2-BM} }
\begin{tabular}{|c|c|c|c|c|c|c|}
\hline 
\textbf{Classifier} & \textbf{Feature Set} & \textbf{Precision} & \textbf{Recall} & \textbf{F1} & \textbf{Accuracy} & \textbf{Weighting Scheme}\tabularnewline
\hline 
 & FS1 & 90.545 & 92.983 & 91.748 & 91.172 & LTC\tabularnewline
\cline{2-7} \cline{3-7} \cline{4-7} \cline{5-7} \cline{6-7} \cline{7-7} 
RF & FS2 & 90.673 & 92.950 & 91.861 & \textbf{91.396} & BM25\tabularnewline
\cline{2-7} \cline{3-7} \cline{4-7} \cline{5-7} \cline{6-7} \cline{7-7} 
 & FS3 & 74.862 & 76.295 & 75.572 & 75.184 & LTC\tabularnewline
\hline 
 & FS1 & 85.643 & 88.737 & 87.163 & 86.868 & LTC\tabularnewline
\cline{2-7} \cline{3-7} \cline{4-7} \cline{5-7} \cline{6-7} \cline{7-7} 
SVM & FS2 & 85.901 & 89.196 & 87.518 & 87.221 & BM25\tabularnewline
\cline{2-7} \cline{3-7} \cline{4-7} \cline{5-7} \cline{6-7} \cline{7-7} 
 & FS3 & 75.951 & 83.590 & 79.588 & 78.375 & LTC\tabularnewline
\hline 
\end{tabular}
\end{table*}

\begin{figure}
\begin{centering}
\includegraphics[width=6in]{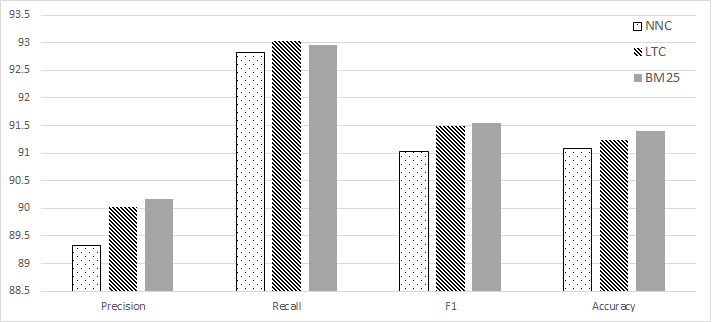}
\par\end{centering}
\caption{Results Comparison of NNC, LTC and BM25 with FS2 on Dataset 2 \label{fig:Result-Comparison-with-dataset2}}
\end{figure}

The result comparison of all three term weighting schemes of FS2 using
RF are visualized in Figure \ref{fig:Result-Comparison-with-dataset2}
of Dataset 2 from which we can justify that $BM25$ weighting scheme
improves precision, f1-score and accuracy. 

\subsection{Results and Analysis on Dataset 3}

\begin{figure}
\begin{centering}
\includegraphics[width=6in]{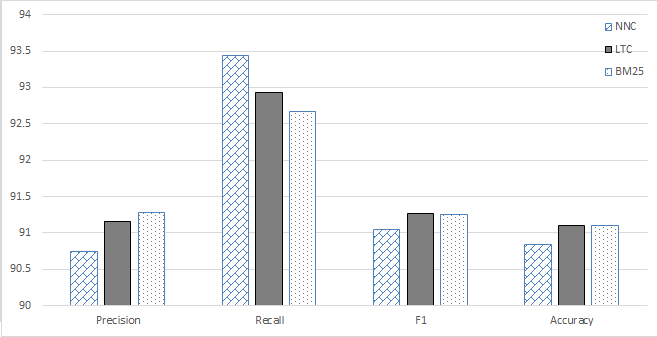}
\par\end{centering}
\caption{Results Comparison of NNC, LTC and BM25 with FS5 on Dataset 3 \label{fig:Result-Comparison-with-dataset3}}
\end{figure}

The comparative results of FS2, FS4 and FS5 based on variation of
RCS on Dataset 3 can be seen in Table \ref{tab:Results-of-Dataset3-BM}.
The feature set FS5 achieved highest precision and accuracy of 91.27\%
and 91.09\% respectively. The improvement of our selected feature
set with BM25 weighting scheme is less than improvement for restaurant
reviews. The less improvement can be justified due to size of hotel
reviews dataset is less than the restaurant reviews dataset. The visualization
of the comparison can be seen in Figure \ref{fig:Result-Comparison-with-dataset3}.
It clearly shows the slight improvement in precision which effects
the overall accuracy of the classification results.

\begin{table}[h]
\caption{Results of Dataset 3 with BM25 \label{tab:Results-of-Dataset3-BM}}

\begin{centering}
\begin{tabular}{|c|c|c|c|c|c|c|}
\hline 
\textbf{Classifier} & \textbf{Feature Set} & \textbf{Precision} & \textbf{Recall} & \textbf{F1} & \textbf{Accuracy} & \textbf{Weighting Scheme}\tabularnewline
\hline 
 & FS4 & 91.226 & 92.782 & 91.102 & 90.967 & LTC\tabularnewline
\cline{2-7} \cline{3-7} \cline{4-7} \cline{5-7} \cline{6-7} \cline{7-7} 
RF & FS2 & 90.453 & 92.910 & 90.943 & 90.838 & BM25\tabularnewline
\cline{2-7} \cline{3-7} \cline{4-7} \cline{5-7} \cline{6-7} \cline{7-7} 
 & FS5 & 91.274 & 92.668 & 91.252 & \textbf{91.096} & BM25\tabularnewline
\hline 
 & FS4 & 85.301 & 89.696 & 87.129 & 86.774 & LTC\tabularnewline
\cline{2-7} \cline{3-7} \cline{4-7} \cline{5-7} \cline{6-7} \cline{7-7} 
SVM & FS2 & 84.936 & 89.821 & 87.117 & 86.709 & BM25\tabularnewline
\cline{2-7} \cline{3-7} \cline{4-7} \cline{5-7} \cline{6-7} \cline{7-7} 
 & FS5 & 85.153 & 89.965 & 87.215 & 86.838 & BM25\tabularnewline
\hline 
\end{tabular}
\par\end{centering}
\end{table}

\section{Performance of Classifiers on Fake Review Detection}

\begin{figure}[H]
\begin{centering}
\includegraphics[scale=0.8]{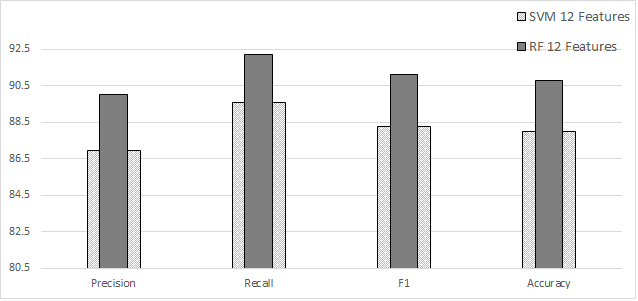}
\par\end{centering}
\caption{Improvement Comparison of RF and SVM with FS2 on Dataset 1 \label{fig:Improvement-Comparison-of}}
\end{figure}

\begin{figure}[H]
\begin{centering}
\includegraphics[scale=0.7]{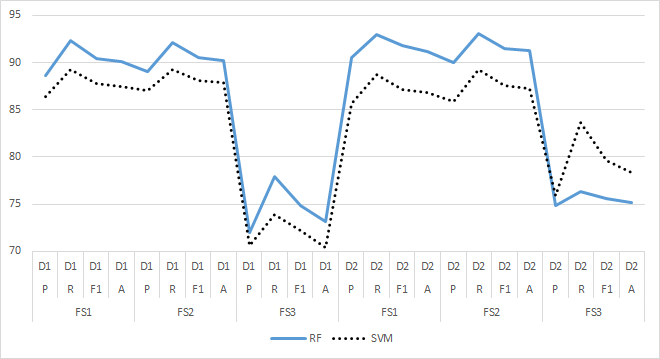}
\par\end{centering}
\caption{Results Comparison of RF and SVM on Dataset 1 and Dataset 2 \label{fig:SVM-RF-All-results}}

\end{figure}

Two classifiers were selected for experimentation which include SVM
and RF. The reason of selecting SVM and RF is that both classifiers
are appreciated in literature for fake review detection. SVM and RF
generated best results on Yelp dataset. We conducted experiments with
three different feature sets. We also computed RCS based on three
different weighting schemes and combined with other behavioral features.
All outcomes of classification supports the statement that RF outperformed
SVM in every evaluation measure. To check the performance of classifiers
we initially conducted the experiments and the results are shown in
Table \ref{tab:Results-dataset1-ltc}. Results show that SVM did not
perform well in comparison with RF. The difference of precision, recall,
f1 and accuracy on FS2 in Dataset 1 is visualized in Figure \ref{fig:Improvement-Comparison-of}.
The improvement of 2.6\%, 3.1\%, 2.2\% and 2.6\% in terms of accuracy,
recall, precision and in f1-measure respectively was observed. Including
the improvement of 4.1\% , 3.97\%, 3.8\% and 4.1\% in terms of in
accuracy, f1-measure, recall and precision with FS2 using $LTC$.
The comparison is shown in Figure \ref{fig:SVM-RF-All-results}. Above
mentioned result tables support that RF outperformed SVM on Dataset
1 and Dataset 2 including all three term weighting schemes. 

\section{Result Analysis}

Experiments conducted with variation of behavioral and contextual
feature sets explored importance of selected features for training
fake review detection model. We compared results of different feature
sets including three different term weighting schemes on SVM and RF.
From initial experiments for exploring importance of \textit{``Review
Deviation}'' with other behavioral and contextual features we analyze
that by new feature improves accuracy. Where as the finding based
on our experimental results shows that by scaling dataset can improve
the classification accuracy and f1-score. Literature on classifier
comparison by \cite{Zhang2016} also reports that RF outperformed
other classifiers. The worthy considerable finding of the conducted
experimentation on variety of feature set and dataset size is that
by adopting term weighting scheme $BM25$ to calculate similarity
of two vectors of reviews can improve the evaluation score. 

\chapter{Conclusion and Future Directions}

\thispagestyle{empty}
\fancyhead[RO]{ CONCLUSION AND FUTURE DIRECTIONS }

This chapter discusses a brief summary of this research work and gives
future direction. It discusses the conclusion drawn on exploring contextual
and behavioral features for fake review detection. Some of the major
contribution is also discussed. It also gives future research direction. 

\section{Summary}

The impact of online user reviews heavily influence the customer decisions
and businesses. Reviews help in decision making of a customer for
purchasing of particular product or service. Fake reviews can mislead
customer in terms of decision making. Fake reviews are use to promote
or demote product or services on e-commerce sites. It can harm the
reputation of a good service or product provider and thus can cause
financial loss a firm. In some cases fake reviews can falsely cause
financial gain to a company or firm. Fake reviews are damaging for
both customers and businesses. Researchers have focused pm fake review
detection since 2007. There are three main research areas: identifying
fake reviews, identifying individual spammer and spammer group. The
focus of this research was to detect fake reviews using contextual
and behavioral features. Researchers have used two types of dataset:
pseudo fake reviews and real life reviews. Using only contextual features
researchers reported high classification results on pseudo fake review
dataset but same classification model failed to acquire high accuracy
on real life review dataset. Our findings reveal some unexplored angle
of fake review detection area for exploitation of behavioral and contextual
features to improve the classification model for identification of
fake reviews in real life dataset.

\section{Conclusion}

In this research, dataset of real life reviews extracted from Yelp
was used. The main focus was to investigate the effect of a behavioral
feature ``reviewer deviation'' with other contextual and behavioral
features on classifying fake reviews. This feature set shows improvement
in classifying fake reviews as compared to the feature sets used by
\cite{Mukherjee2013} and \cite{Zhang2016}. Another focus was to
explore a contextual feature ``reviewer content similarity'' (RCS)
using different weightings. Researchers have used RCS based on TF.IDF
\cite{Jindal2007,Zhang2016,algur2010spam,Banerjee2014,LinY2014,Ott2013}.
RCS based on NNC and BM25 along with other contextual and behavioral
features shows improvement in real life review dataset. Our findings
reveal some unexplored angle of fake review detection area. It was
observed that RF outperforms SVM in detecting fake reviews on real
life dataset and this fact is reported by several researchers \cite{Zhang2016,Mukherjee2013}. 

\section{Contributions}

A set of predictive features containing behavioral and contextual
features to identify fake reviews was investigated. A behavioral feature
``reviewer deviation'' was not explored in any of the feature set
used to identify fake reviews. The perspective of ``reviewer deviation''
is to capture different behavior of reviewer by calculating deviation
of given ratings with other ratings on same restaurant/hotel. The
proposed idea to combine ``reviewer deviation'' with other contextual
and behavioral feature for fake review classification shows improvement.
The importance of ``reviewer deviation'' with other predictive features
was investigated. The result shows that ``reviewer deviation'' is
among top ten most important feature in terms of importance score.
Another contribution was to investigate a contextual feature ``reviewer
content similarity'' (RCS) based on different weightings: NNC and
BM25. Results on RCS based on NNC and BM25 shows improvement in classifying
fake reviews. 

\section{Future Work}

There are three research areas associated with fake review detection:
identifying fake reviews, identifying individual spammer and spammer
group. Identifying group spammer can be explored in future using social
network analysis techniques. Further feature selection can be pruned
using deep learning techniques. RCS can be explored further using
text processing techniques.

\newpage{}

\renewcommand\bibname{References}
\fancyhead[RO]{ }
\fancyhead[LO]{ }

\bibliographystyle{plain}
\bibliography{main}

\end{document}

%% file: front.tex
\begin{titlepage}
	\begin{center}
		{
			\Huge
			\textbf{\newline Fake Review Detection Using Behavioral and Contextual Features}\\
		}
		
		\vspace{0.3cm}
		\begin{figure}[!h]
			\centering
			\includegraphics[scale=0.2]{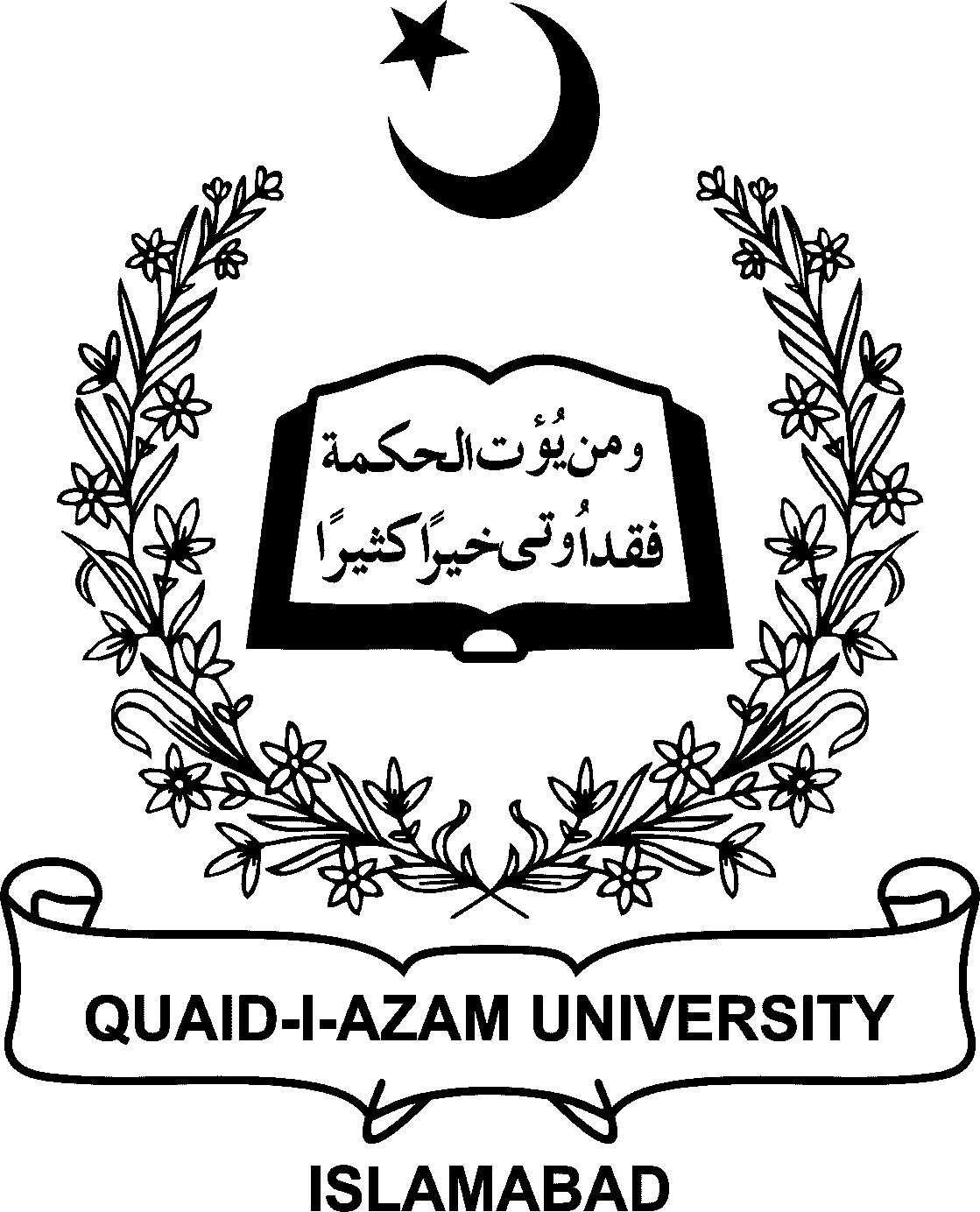}
			\label{fig:Logo1}
		\end{figure}
		
		\vspace{8mm}
		
		{
			\Large
			By\\[0.2\baselineskip]
			\vspace{2mm}
			%			\textsc
			{Jay Kumar}
		}
		
		\vspace{6mm}
		{
			
			\LARGE
			\textbf{Department of Computer Science}\\
			\textbf{Quaid-i-Azam University}\\				
			\textbf{Islamabad, Pakistan}\\			
			\vspace{0.8cm}		
			\textbf{February, 2018}
		}
		
	\end{center}
\end{titlepage}

% ///////////////////////////////// BIGIN COMMENTS /////////////////////////////////////////%
%\begin{comment}

\begin{titlepage}
	\begin{center}
		
		{
			\Huge
			\textbf{\newline Fake Review Detection Using Behavioral and Contextual Features}\\
		}
		
		\vspace{0.1cm}
		\begin{figure}[!h]
			\centering
			\includegraphics[scale=0.18]{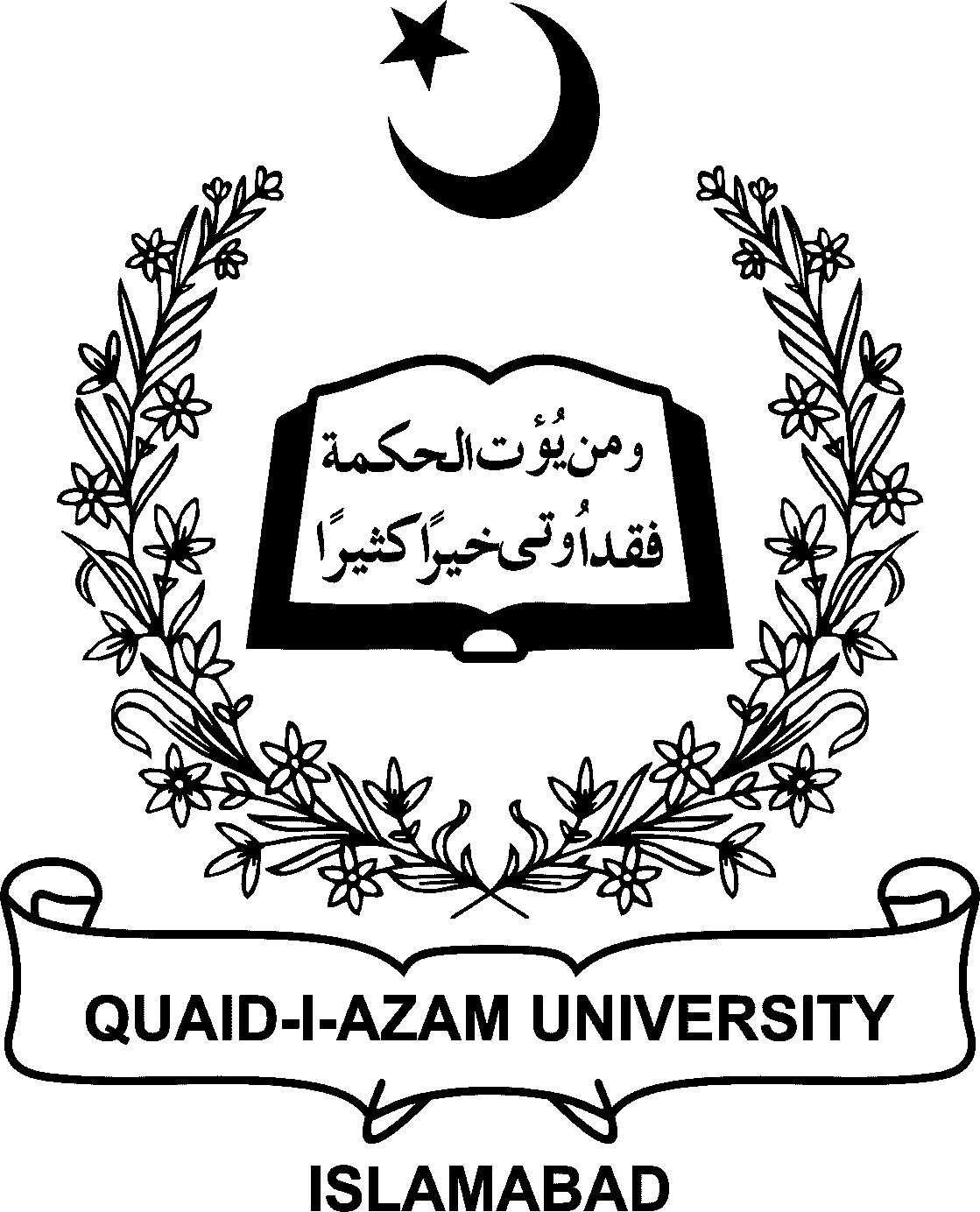}
			\label{fig:logo2}
		\end{figure}
		
		\vspace{0.1mm}
		
		{
			\Large
			By\\[0.1\baselineskip]
			\vspace{0.1mm}
			%			\textsc
			{Jay Kumar}
			\\
			\vspace{1mm}
			\textbf{Supervised by}\\
			\vspace{0.2mm}
			Dr. Akmal Saeed Khattak\\
		}
		\vspace{0.2mm}
		
		{
			\LARGE
			\textbf{Department of Computer Science}\\
			\textbf{Quaid-i-Azam University}\\		
			\textbf{Islamabad, Pakistan}\\		
			\vspace{0.4cm}	
			\textbf{February, 2018}
		}
		
	\end{center}
\end{titlepage}
\begin{titlepage}
	\begin{center}

		{
			\huge
			\textbf{\newline Fake Review Detection Using Behavioral and Contextual Features}\\
		}

		\vspace{0.001cm}
		\begin{figure}[!h]
			\centering
			\includegraphics[scale=0.13]{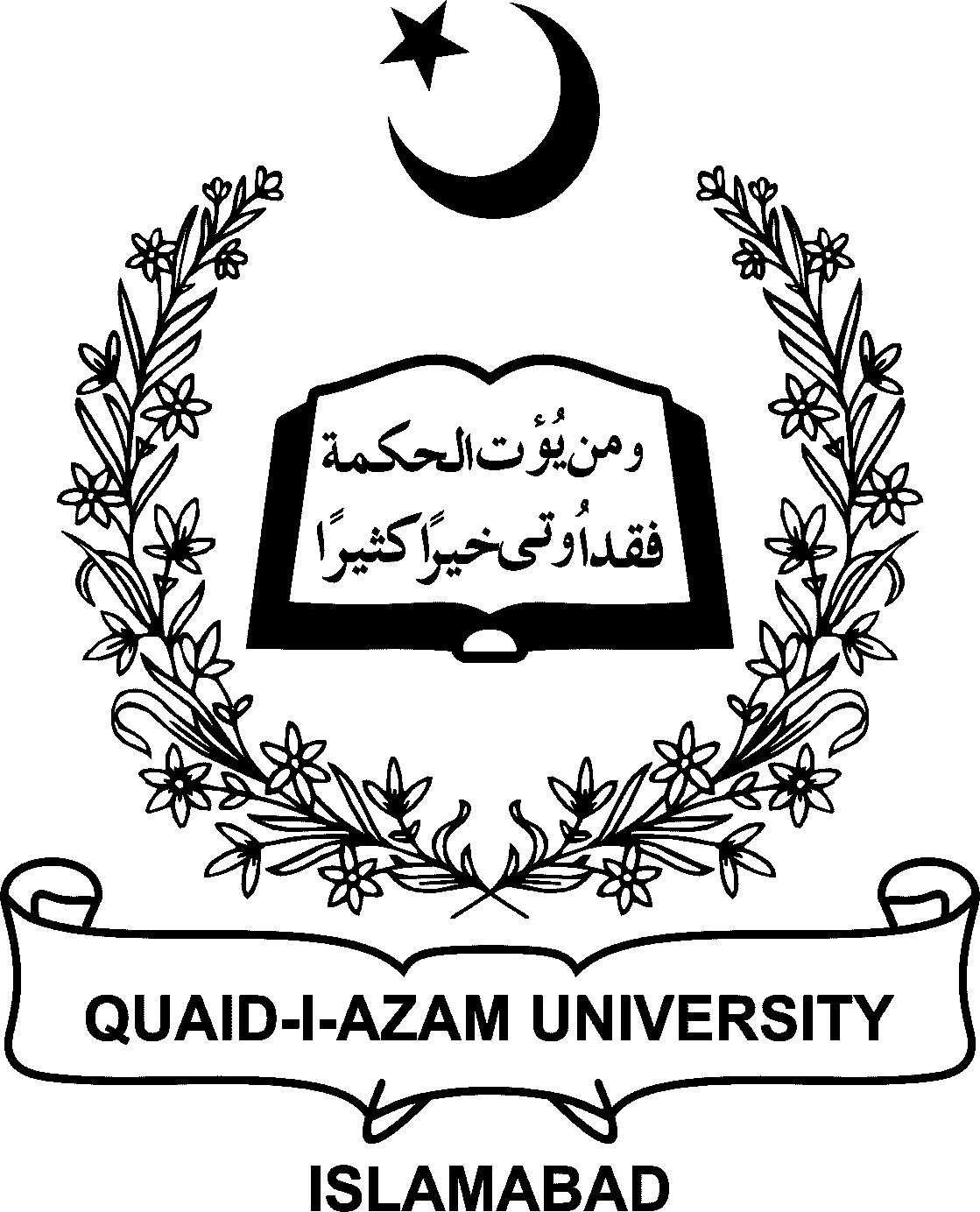}
			\label{fig:logo3}
		\end{figure}
		
		{
			\Large
			By\\[0.0001\baselineskip]
			
			{Jay Kumar}\\[0.0001mm]
		}
		{
			\large
			\textit{A Dissertation Submitted in Partial Fulfillment for the}\\
			\textit{Degree of}\\
			
			{MASTER OF PHILOSOPHY}\\			
			
			IN\\
			COMPUTER SCIENCE\\
			
		}
		{
			\LARGE
			\textbf{\bf Department of Computer Science}\\
			\textbf{\bf Quaid-i-Azam University}\\
			\textbf{Islamabad, Pakistan}\\			
			\textbf{February, 2018}
		}
		
	\end{center}
\end{titlepage}

\def\dedication{
	\Large DEDICATION \\
	
	\textit{ \centering  To my beloved mother }\\
	\textit{ \centering  To my beloved father }\\
	\textsl{ \centering  To my beloved brother }\\
	\textsl{ \centering  To my best friends }\\
}

\def\acknowledgementtext{
	Praise is to \textbf{God}, the Almighty who have been giving everything to me and guiding me at every stage of my life. Bundle of thanks and gratefulness go to my supervisor \textbf{Dr. Akmal Saeed Khattak}, without his guidance and follow-up; this research would never have been. In addition, I would like to extend my thanks to Dr. Khalid Saleem, who supported me during my degree period. Additionaly, I want to thank Dr. Onaiza Maqbool, Dr. Ghazanfar Farooq, Dr. Muddassar Azam Sindhu, and Dr. Shuaib Kareem who shared useful knowledge in different courses. I would like to thank Dr. Muhammad Usman and non-academic staff who provided all needed facilities and support. I would also like to thank Dr. Bing Liu for providing dataset.
	
	I would like to thank  Maryam Javed, Nouman Khan, Naveed Tariq, Aqib Rehman, Muhammad Awais Amjad, Farooq Zaman, Anjum Ara, Muhammad Usman, Qamar Munir, Muhammad Islam, Zaffar Saeed, Khwaja Bilal, Umer Naeem, Younis Iqbal, Aftab Rashid, classmates and junior fellows for making my study a great experience, motivational, useful, enjoyable, and full of joyful atmosphere. Last but not least, I am greatly indebted to my \textbf{mother, father and brother} for motivation, encouragement and always supporting me on every part of life. \\

	\textit{Jay Kumar}
}

\def\abstracttext{
User reviews reflect significant value of product in the world of e-market. Many firms or product providers hire spammers for misleading new customers by posting spam reviews. There are three types of fake reviews, untruthful reviews, brand reviews and non-reviews. All three types mislead the new customers. A multinomial organization "Yelp" is separating fake reviews from non-fake reviews since last decade. However, there are many e-commerce sites which do not filter fake and non-fake reviews separately. Automatic fake review detection is focused by researcher for last ten years. Many  approaches and feature set are proposed for improving classification model of fake review detection. There are two types of dataset commonly used in this research area: psuedo fake and real life reviews. Literature reports low performance of classification model real life dataset if compared with pseudo fake reviews. After investigation behavioral and contextual features are proved important for fake review detection\\
Our research has exploited important behavioral feature of reviewer named as "\textit{reviewer deviation}". Our study comprises of investigating \textit{reviewer deviation} with other contextual and behavioral features. We empirically proved importance of selected feature set for classification model to identify fake reviews. We ranked features in selected feature set where \textit{reviewer deviation} achieved ninth rank. To assess the viability of selected feature set we scaled dataset and concluded that scaling dataset can improve recall as well as accuracy. Our selected feature set contains a contextual feature which capture text similarity between reviews of a reviewer. We experimented on NNC, LTC and BM25 term weighting schemes for calculating text similarity of reviews. We report that BM25 outperformed other term weighting scheme.\\

}
\def\declaration{
I hereby declare that this dissertation is the presentation of my original research work. Wherever contributions of others are involved, every effort is made to indicate this clearly with due reference to the literature and acknowledgment of collaborative research and discussions.

This work was done under the guidance of Dr. Akmal Saeed Khattak , Department of Computer Sciences, Quaid-i-Azam University, Islamabad (Pakistan).
	
\begin{table}[h]
	\centering
			\begin{tabular}{lp{2.5in}p{1.4in}}
			\tabularnewline
			\tabularnewline
			\tabularnewline
			\tabularnewline
				Date: \underline{23rd February, 2018}  &&\\
				&&\rule{\linewidth}{0.1mm}\\
				&& Jay Kumar  \\
		\end{tabular}
		
\end{table}
 }

\newpage
\vspace*{2in}

\begin{center}{\dedication}\end{center}
\pagenumbering{roman}

\chapter*{Declaration}
\declaration
\clearpage

\chapter*{Abstract}
\addcontentsline{toc}{chapter}{Abstract}
\abstracttext
\clearpage

\chapter*{Acknowledgment}
\acknowledgementtext
\clearpage

\tableofcontents
\thispagestyle{empty}
\clearpage

\listoftables
\addcontentsline{toc}{chapter}{List of Tables}
\clearpage

\listoffigures
\addcontentsline{toc}{chapter}{List of Figures}
\clearpage